\documentclass{article}
\usepackage{amsmath, amssymb, graphics}

\newcommand{\mathsym}[1]{{}}

\begin{document}

\title{  { } { } { } { } { } { } { } { } { }Introduction to Quantum Message Space}
\author{\\
Technical Report $\#$2, Quantum Computation Study Group, { }\\
Department of Computer Science and Engineering,\\
\hspace*{0.5ex} Texas State University { }at San Marcos\\
\hspace*{0.5ex} \\
\hspace*{0.5ex} Prepared by R. D. Ogden\\
ro01@txstate.edu\\
Revised November 2, 2005}
\date{}
\maketitle

\begin{abstract}\textrm{ \\
\hspace*{3.5ex} According to Landauer's Principle the annihilation (and presumably the creation) of one bit costs at least }\textit{ kT} ln(2)\textrm{
 much energy. This seems to imply { }that serial bit transmission is impossible; { }however, a complete quantum theory { }of communication should
permit Alice to send Bob a message of arbitrary length, unknown to Bob in advance. { }One is lead to consider the Hilbert state space { }}\(\left.\left.\sum
_{n=0}^{\infty } \right(\mathbb{C}^2 \right)^{\otimes \text{\textit{$n$}}}\)\textrm{ , where }\(\mathbb{C}^2\)\textrm{  is conventional qubit space,
but the obvious operator which appends a 0 (say) to a string }x (i.e., sends $\mid $\textit{ x}$\rangle $ to  $\mid $\textit{ x0}$\rangle $ ) is
no longer unitary on this space because it is not onto\textrm{ . The remedy proposed in this report is to use { }}\(\ell ^2\)\textit{ (FG}(2)) \textrm{
for the quantum message space (QMS), where }\textrm{ \textit{ FG}}\textrm{ (2) is the free group on the bit symbols $\{$0,1$\}$; the $\texttt{"}$anti-bits$\texttt{"}$
}\(0^{-1}\) and \(1^{-1}\), represented by 2 and 3 respectively,\textrm{  are introduced, and thus the conservation laws are retained. After developing
the basic QMS constructs we describe the message-length observable }\textrm{ \textit{ N}}\textrm{ , noting that most quantum operators used hitherto
in quantum information theory { }commute with }\textrm{ \textit{ N}}\textrm{ . In QMS the operation which appends a string to a message is implemented
by a right translation { }operator. We review the harmonic analysis on the free group }\textrm{ \textit{ FG}}\textrm{ (2) { }and the decomposition
of { }the right regular representation into irreducible representations of }\textrm{ \textit{ FG}}\textrm{ (2). This decomposition is implemented
by the spectral analysis of the operator }\textrm{ \textit{ A}}\textrm{ , { }which is left convolution by}\textrm{  }($\mid $\textit{ 0}$\rangle
$+$\mid $\textit{ 1}$\rangle $+$\mid $\textit{ 2}$\rangle $+$\mid $\textit{ 3}$\rangle $)/\textrm{ 4}\textrm{  }\textrm{ . This analysis yields a
family of projectors which commute with the extended qubit operations implemented by the right regular representation and therefore can be used to
construct quantum operators for quantum computation on QMS.\\
 { } { } { } }
\end{abstract}

\section*{\\
I. { } { }INTRODUCTION}

In classical communication theory a binary-encoded message source generates bit strings according to the underlying source statistics, those of an
ergodic stochastic process. In general, messages of arbitrary length may be generated. { } { }Moreover, in classical communication theory it is convenient
to conceive of the bits in a message being generated serially one at a time.\\
In much of quantum computation theory the message length is assumed known; i.e., \textit{ N}-qubits are studied with \textit{ N}, the number of qubits,
fixed. The underlying Hilbert space is then a \(2^N\)\textit{  }dimensional complex space with a complete orthogonal set of normalized pure states
indexed by the strings of length \textit{ N}. Furthermore, it is often assumed that the {``}noise{''} in the quantum communication channel preserves
\textit{ N} , but it has been argued that there is no physical { } { }necessity for this.\\
In contrast, quantum message space permits strings of arbitrary length; it contains an orthonormal list \textit{ (}\(\delta _a\)\textit{ ) }of orthonormal
state vectors indexed by { } { }\textit{ a} $\in $ 2*  { } { }= the set of \textit{ all} bit strings. The message is then an observable whose value
\textit{ a} is a (possibly empty) { }bit string. If $\varrho $ is the \textit{ a priori} state before measurement, the probability that the message
received equals \textit{ a}\textup{  }\textup{ is}\textup{  }\( \left\langle \delta _a\mid \varrho \left(\delta _a\right)\right\rangle /\text{Trace}(\varrho
\text{\textit{$)$}}\text{\textit{$ $}}\)\textit{ .} { } { }When a measurement is made with result \textit{ a} (i.e., a { } { }quantum message whose
value is \textit{ a} is received), the system state becomes $\langle $\(\delta _a\)$\mid $$\varrho $(\(\delta _a\))$\rangle $ \(\delta _a\)$\otimes
$\(\delta _a^*\), where the { } { }operator $\alpha $$\otimes $\(\beta ^*\) is defined by $\alpha $$\otimes $\(\beta ^*\)($\varphi $)=$\langle \beta
\mid \varphi \rangle $ $\alpha $ . Let \(\mathcal{H}^+\) = Hilbert space generated by { } { }\(\left(\delta _a\right)_{a \in 2^*}\) { } { }. A state
$\varrho $ supported by \(\mathcal{H}^+\) represents a $\texttt{"}$positive$\texttt{"}$ message source; i.e., one whose value is a (possibly empty)
{ }string of \textit{ 0}s and \textit{ 1}s.\\
The space \(\mathcal{H}^+\) would seem to be a natural candidate for a quantum state space which allows messages of arbitrary length, but it leaves
no provision for bit creation.\textit{  }The obvious process of bit creation by appending is implemented by { }the operator \textit{ V}\(_b\) which
maps \(\delta _a\) to \(\delta _{a\text{\textit{$b$}}}\) ,\textit{  }where \textit{ b} $\in $ 2 { }is a bit; { } { }e.g., { } \(V_0\)(\(\delta _{001}\))=\(\delta
_{0010}\) . { }\textit{ V}\(_b\) leaves \(\mathcal{H}^+\) invariant and acts as an isometry thereon, but this map is not onto , thence it is not
unitary.\textit{  }This is consistent with Landauer's principle[NG]: { }since at least\pmb{  }\textit{ kT ln(2) } much  energy is required to erase
a bit, presumably at least that much energy is needed to create a new bit. Thus it would seem that serial bit generation is not possible in a closed
system.

The operations of appending and deleting bits evoke the creation operators (e.g., \(V_b\)) and annihilation operators  (\(V_b^*\)) of quantum field
theory. The formal analogues of momentum and position would be the real and imaginary parts of { }\(V_b\) . { }The empty string is sort of like the
vacuum from which everything is created. But the (anti-)commutation relations are disappointing and trivial; with hindsight this is so because \(2^*\)
is a free semi-group.

But there is a precedent in atomic physics for evading conservation laws by postulating a new {``}particle{''} with the necessary properties for
balancing the equations; in such a way the positron and neutrino were first proposed, by Pauli/Dirac and E. Fermi, and sure enough physical { }evidence
turned up that seemed to confirm their reality, which today { }is an established fact of physics. In this paper anti-qubits are proposed, which do
make the appending of a bit a unitary operation, balanced by effects on anti-qubits and mixtures. But in this case we are ignorant of any physical
basis in new particles or fields, but of new \textit{ relations }within infinite-dimensional Hilbert spaces suitable for modelling a quantum message
space which permits bit generation and messages of arbitrary length unknown in advance.\\
The basic references for quantum mechanics used in this report are Mackey, $\texttt{"}$Mathematical Foundations of Quantum Mechanics$\texttt{"}$
[Ma] and the standard reference of Nielsen and Chaung, $\texttt{"}$Quantum Computation and Quantum Information$\texttt{"}$ [NG]. { }[PS] is a suggested
reference for quantum field theory.\\
Special thanks to Carl Brannen for a close critical reading and various helpful suggestions.

\section*{\\
II. THE FREE GROUP FG(2): { } { }ENCODINGS AND CONVENTIONS}

\subsection*{\\
A. Strings}

\(2^n\) refers to the set of integers  $\{$0,1,2,3,...,\(2^n\)-1$\}$, or the set of bit strings of length \textit{ n}, or the simply the number \(2^n\),
depending on the context. The set of all bit strings { }\(2^*\) = \(\cup _{n\geq 0}\) \(2^n\) forms a semigroup with identity under the operation
of string concatenation.\\
When we complete \(2^{* }\) as a group we get the free group with two generators, \textit{ FG(2)} . We will use \textit{ 0} and \textit{ 1} for the
group generators, and denote \(\text{\textit{$0$}}^{\text{\textit{$-1$}}}\) and \(\text{\textit{$1$}}^{\text{\textit{$-1$}}}\) { } { }by \textit{
2} and \textit{ 3} respectively. The set of generators and their inverses is the alphabet for the free group, so for\textit{  FG(2)} this shall be
4 = $\{$0,1,2,3$\}$ . 0 and 1 are bits that make up positive messages; 2 and 3 { } { }(in this context) are the \textit{ anti-bit} values. The free
group identity is the empty string { } { }$\Lambda $ . The inverse map on 4 is the map \textit{ inv: }0 $\rightarrow $ 2, 1$\rightarrow $ 3, 2 $\rightarrow
$ 0, 3 $\rightarrow $1, \textit{ a}$\rightarrow $\textit{  }\(a^{-1}\).\\
In a free group there are no simplifications possible except that of removing digrams consisting of an alphabet symbol and its inverse: { }\textit{
a}\(a^{-1}\)\textit{  .} In the present context this means that any of the four substrings \textit{ 02,13,20,31} may be removed from an expression
representing a free group element. A string in \(4^*\) which does not contain one of these four substrings is a \textit{ reduced word. }

\subsection*{\\
B. The Free Group}

The free group on two elements \textit{ FG(2)} is the subset of \(4^*\) consisting of all reduced words. The group operation is string concatenation
followed by reduction, \textit{ $\Lambda $} is the identity element and the inverse of the reduced word is the string of the inverse alphabet symbols,
in reverse order. { } { }\textit{ \\
}\(\text{\textit{FG}}(2)_n\) = \textit{ FG}\textup{ (}\textup{ 2}\textup{ )}$\cap $ \(4^n\)\textit{  = }all reduced words of length \textit{ n. }The
group operations on \textit{ FG(2)} are simply \\
 { } { }\textit{ x $\cdot $ y}\textit{  }\textit{ =}\textit{  }\textit{ reduce}\textit{ (x y)} . If { } { }\textit{ x}\textit{  }=\(x_0\)\textit{
...}\(x_{n-1}\) then \(x^{-1}\)=inv(\(x_{n-1}\))...inv(\(x_0\))\textit{  }\textit{ .}\\
Example: 03221211$\cdot $3300123=032210123=\((103230012)^{-1}\) { } { } \pmb{ .}\\
The mathematical significance of the free group \textit{ FG(2)} is given any group G and pair of elements \textit{ a} and \textit{ b}, then there
is a unique homomorphism from \textit{ FG(2)} onto the subgroup of G generated by\textit{  $\{$a,b$\}$} which sends 0 to \textit{ a} and 1 to \textit{
b}. { }The significance for quantum communication theory may be due to its being the natural mathematical structure which allows for bit generation
as a unitary operation. { } { }\textit{ \\
}Note we have the inclusions \(2^n\)$\subset $\(\text{\textit{FG}}(2)_n\) $\subset $\(4^n\) { } { }.

\subsection*{C. Definition Of Quantum Message Space (QMS)}

A quantum message space is a Hilbert space { } { }$\mathcal{H}$ together with an orthonormal basis \(\left(\delta _a\right)_{a \in \text{\textit{FG}}\text{\textit{$($}}\text{\textit{$2$}}\text{\textit{$)$}}}\)
indexed by the elements of the free group \textit{ FG}(2) on the two bit values $\{$0,1$\}$. The basis induces a unique atomic spectral measure \textit{
M} mapping subsets \textit{ A $\subseteq $ }\textit{ FG}(2) to orthogonal projectors:

\begin{equation} \text{\textit{$M(A)$}} = \sum _{a\in A}\text{    }\delta _a\otimes \delta _a^*,\text{    }A\subseteq \text{FG}(2) \end{equation}

The elements of \textit{ FG}(2), the reduced strings, are the outcome of measurements of the discrete observable determined by \textit{ M}. Thus
the measured message is a reduced substring on $\{$0,1,2,3$\}$ . Strings of 0s and 1s are reduced; such strings might be called \textit{ positive
}messages.\\
Let ($\varphi $,$\psi $) $\to $ $\langle \varphi \mid \psi \rangle $ denote the inner product on $\mathcal{H}$, anti-linear in $\varphi $. Any quantum
state $\varrho $ in message space can be interpreted as a message source. It induces a probability distribution on the messages,\textit{  x $\in
$ }\textit{ FG(2)} having weight $\langle $\(\delta _{\text{\textit{$x$}}}\)$\mid $$\varrho $(\(\delta _{\text{\textit{$x$}}}\))$\rangle $/Trace($\varrho
$)=probability that the message $\texttt{"}$received$\texttt{"}$ is \textit{ x}. If the outcome of the measurement is message \textit{ x} then the
\textit{ a posteriori} state is the pure state $\langle $\(\delta _{\text{\textit{$x$}}}\)$\mid $$\varrho $(\(\delta _{\text{\textit{$x$}}}\))$\rangle
$\(\delta _{\text{\textit{$x$}}}\)$\otimes $\(\delta _{\text{\textit{$x$}}}^*\) .\\
\hspace*{0.5ex} The \textit{ message source entropy} in the quantum message space is

\[\text{        }\sum _{x\in \text{\textit{$\text{FG}(2)$}}}\eta (\left\langle \delta _{\text{\textit{$x$}}}\mid \varrho \left(\delta _{\text{\textit{$x$}}}\right)\right\rangle
/\text{Trace}(\varrho )), \text{where}\text{  }\eta \text{\textit{$(p)$}} = \text{\textit{$-p \ln (p)$}}\text{\textit{$,$}}\text{\textit{$ $}}0<p\leq
1,\text{\textit{$ $}}\eta (0) = 0\]

This entropy depends on the choice of the particular basis (\(\delta _a\)) ; it is not the same as the von Neumann entropy = Trace($\eta $($\varrho
$)), but as shown in appendix A the von Neumann entropy is a lower bound for the source entropy with respect to (\(\delta _a\)).

\subsubsection*{ { } { } { } { }1) { } { } { } { } \pmb{  { } { }Message { }Length Observable:}}

\(\text{    }\text{The}\text{    }\text{message}\text{    }\text{length}\text{    }\#x\text{    }\text{is}\text{    }\text{an}\text{    }\text{important}\text{
   }\text{observable}\text{  }\text{which}\text{    }\text{we}\text{    }\text{denote}\text{    }\\
\text{by}\text{    }\text{\textit{$N$}} .\text{    }\text{Formally},\text{we}\text{    } \text{equate}\text{    }\text{\textit{$N$}}\text{\textit{
   }}\text{with}\text{    }\text{the}\text{    }\text{unbounded}\text{    }\text{self-adjoint}\text{    }\text{operator} \)

\begin{equation}\text{                    }\text{\textit{$N$}}\text{\textit{$ $}}\text{\textit{$=$}}\text{\textit{$ $}}\sum _{n=0}^{\infty } \text{\textit{$n$}}\text{\textit{$
$}}\text{\textit{$M$}}\left(\text{\textit{FG}}(2)_n\right)\end{equation}

\(\text{The}\text{    }\text{\textit{$ $}}\text{\textit{average}}\text{\textit{    }}\text{\textit{message}}\text{\textit{    }}\text{\textit{length}}\text{\textit{
   }} \text{of}\text{     }\text{the}\text{    }\text{quantum}\text{    }\text{message}\text{    }\text{source}\text{    }\varrho \text{    } \text{is}\)

\(\sum _{x\in \text{FG}(2)} \left.\#x\text{  }\left\langle \delta _{\text{\textit{$x$}}}|\varrho \left(\delta _{\text{\textit{$x$}}}\right)\right\rangle
\right/\text{Trace}(\varrho )\text{  }= \text{Trace} (\text{\textit{N$\varrho $}}\text{\textit{$)$}}, \text{which}\text{    }\text{may}\text{   
}\text{be}\text{    }\text{infinite}.\)

\( \text{Note}\text{    }\text{that}\text{  }\#\text{\textit{FG}}(2)_0=1, \text{and}\text{  }\text{for}\text{  }\text{\textit{$n$}}\text{\textit{$>$}}\text{\textit{$0$}}\text{\textit{$,$}}\text{\textit{$
$}}\text{  }\#\text{\textit{FG}}(2)_n\text{ = }4\cdot 3^{n-1} .\)

\(\text{Thus}\text{  }\text{the}\text{  }\text{positive}\text{  }\text{semi-definite}\text{  }\text{operator}\text{  }\text{\textit{$N$}}\text{\textit{$
$}}  \text{has}\text{  }\text{the}\text{  }\text{property}\text{   }\text{that}\text{  }\text{the}\text{  }\\
\text{multiplicity}\text{  }\text{of}\text{  }\text{an}\text{  }\text{eigenvalue}\text{  }\text{grows}\text{  }\text{exponentially}\text{  }\text{with}\text{
 }t \text{he}\text{  }\text{size}\text{  }\text{of}\text{  }\text{the}\text{  }\text{eigenvalue} .\)

\(\text{After}\text{    }a\text{    }\text{measurement}\text{    }\text{of}\text{    }\text{\textit{$N$}}\text{\textit{    }}(\text{but}\text{  
 }\text{not}\text{    }\text{of}\text{    }\text{the}\text{    }\text{message}\text{    }\text{itself})\text{     }\text{the}\text{    }\text{state}\text{
   }\text{becomes}\\
\text{    }\text{\textit{$M$}}\text{\textit{$($}}\text{\textit{FG}}\text{\textit{$(2)$}}_n\text{\textit{$)$}}\varrho \text{\textit{$ $}}M\left(\text{\textit{FG}}(2)_n\right.),\text{
   }\\
\text{where}\text{    }\text{\textit{$n$}}\text{\textit{    }} \text{is}\text{    }\text{the}\text{    }\text{result}\text{    }\text{of}\text{ 
  }\text{the}\text{    }\text{measurement}.\text{    } \text{In}\text{    }\text{the}\text{    }\text{\textit{$a$}}\text{\textit{$ $}}\text{\textit{posteriori}}\text{\textit{
 }}\text{state} \\
\text{the}\text{    }\text{observable}\text{    }\text{\textit{$N$}}\text{    }\text{has}\text{    }\text{the}\text{    }\text{definite}\text{  
 }\text{value}\text{    }\text{\textit{$n$}}.\text{    }A\text{    }\text{state}\text{     }\vartheta \text{     }\text{where} \text{in}\text{  
 }\text{the}\text{    }\\
\text{message}\text{    }\text{has}\text{    }\text{definite}\text{     }\text{length}\text{    }\text{\textit{$n$}}\text{     }\text{has}\text{
   }\text{the}\text{    }\text{property}\text{    }\text{that}\text{     }\vartheta  N = n \vartheta ,\text{    }\\
\text{and}\text{    }\text{therefore}\text{    }N \vartheta  = \vartheta  N = \vartheta  N \vartheta  .\text{    }\text{It}\text{    }\text{follows}\text{
   }\text{that}\)

\begin{equation}\text{    }\text{         }\vartheta  = \sum _{x,y\in \text{\textit{FG}}(2)_n}\left\langle \delta _{\text{\textit{$x$}}}|\vartheta
\left(\delta _y\right)\right\rangle  \delta _x\otimes \delta _y^* . \end{equation}

\[\text{If}\text{    }\vartheta \text{    }\text{is}\text{    }\text{also}\text{  }a\text{  }\text{positive}\text{     }\text{message}\text{    
}\text{state},\text{    }\text{                            }\vartheta  = \sum _{x,y\in 2^n}\left\langle \delta _{\text{\textit{$x$}}}|\vartheta \left(\delta
_y\right)\right\rangle  \delta _x\otimes \delta _y^* .\text{                         }\]

\(\text{In}\text{    }\text{particular},\text{    }\text{an}\text{    }\text{\textit{extended}}\text{\textit{    }}\text{\textit{n-qubit}}\text{\textit{
    }}\text{is}\text{    }a\text{    }\text{pure}\text{    }\text{state}  \text{of}\text{    }\text{the}\text{    }\\
\text{form}\text{    }\psi \otimes \psi ^* in\text{    }\text{which}\text{    }\text{\textit{$N$}}\text{\textit{    }}\text{has}\text{    }\text{the}\text{
   }\text{definite}\text{    }\text{value}\text{    }\text{\textit{$n:\text{    }N(\psi )$}}\text{\textit{$=$}}\text{\textit{n$\psi $}}.\\
A\text{    }\text{conventional}\text{     }\text{\textit{n-}}\text{qubit}\text{    }\text{is}\text{    }a\text{    }(\text{normally}\text{    }\text{normalized})\text{
   }\text{pure}\text{    }\text{state}\text{  }\text{in}\text{  }\text{which}\\
\text{  }\psi  \in  \text{Span}\left\{\delta _x:\text{  }x \in  2^n \right\}\text{  }.\text{  }\text{Such}\text{    }a\text{    }\text{state}\text{
    }\text{represents}\text{    }a\text{    }\text{positive}\text{    }\text{\textit{n-}}\text{bit}\text{    }\text{message}. \)

\(\text{It}\text{   }i s\text{  }\text{interesting}\text{   }\text{to}\text{  }\text{explore}\text{  }\text{the}\text{  }\text{dynamics}\text{  }\text{of}\text{
 }a\text{  }\text{quantum}\text{  }\text{system}\text{  }\text{with}\text{  }\\
\text{Hamiltonian}\text{  }\text{\textit{$H$}}\text{ = }\text{\textit{$h$}}\nu \text{\textit{$N$}} .\text{  }\text{Here}\text{  }\text{\textit{$h$}}\text{
 }\text{is}\text{  }\text{Planck}'s\text{  }\text{constant}\text{  }\text{and}\text{  }\nu >0 \text{  }\text{is}\text{  }\text{some}\text{  }\\
\text{positive}\text{  }\text{constant}\text{  }\text{frequency} .\text{  }T \text{he}\text{  }\text{energy}\text{  }\text{eigenvalues}\text{  }\text{are}\text{
  }\text{\textit{$n$}}\text{\textit{$ $}}\text{\textit{$h$}}\nu \text{  }\text{for} n=0,1,\\
2,\ldots  \text{and}\text{  }\text{for}\text{  }n>0\text{  }\text{the}\text{  }\text{eigenvalue}\text{  }\text{\textit{$n$}}\text{\textit{$ $}}\text{\textit{$h$}}\nu
\text{  }\text{has}\text{  }\text{multiplicity}\text{   }\#\text{FG}(2)_n= 4\cdot 3^{n-1}.\)

\(\text{The}\text{  }\text{dynamical}\text{  }\text{unitary}\text{  }\text{operator}\text{   }U_t\text{  }= \text{\textit{$\exp $}}\text{\textit{$($}}\text{\textit{$-2$}}\pi
 i\text{  }N \nu \text{\textit{$t$}}\text{\textit{$)$}}\text{\textit{$ $}} \text{  }\text{has}\text{  }\text{period}\text{  }1/\nu  . \\
 \text{Let}\text{  } \psi  = \sum _{n=0}^{\infty } \psi _n, \text{  }\text{where}\text{    }\psi _n= M\left(\text{\textit{FG}}(2)_n\right.)(\psi
)\text{  } \text{is}\text{  }\text{the}\text{  }\text{orthogonal}\text{  }\text{projection}\text{  }\text{of}\text{  }\psi \text{   }\text{to}\text{
 }\\
\text{the}\text{   }\text{space}\text{   }\text{of}\text{  }\text{messages}\text{  }\text{of}\text{  }\text{length}\text{  }\text{\textit{$n$}}\text{\textit{$.$}}\text{\textit{
 }}\text{Then}\text{  }\text{the}\text{  }\text{time}\text{  }\text{evolution}\text{  }\text{is}\text{  }\text{given}\text{   }\text{by}\text{  }a\text{
 }\text{Fourier}\text{  }\text{series}\)

\begin{equation}\text{                }U_t(\psi ) = \sum _{n=0}^{\infty } \psi _ne^{-2\text{$\pi $i}\text{\textit{$n$}}\nu \text{\textit{$t$}}}\text{\textit{$,$}}\text{\textit{$
$}}\text{\textit{$$}}\text{\textit{                                 }}U_t(\psi )\otimes U_t(\psi )^* = \sum _{n=0}^{\infty } \psi _n\otimes \psi
_n^* + \sum _{m<n} 2\text{Re}(e^{2\text{$\pi $i}(m-\text{\textit{$n$}}\text{\textit{$)$}}\nu \text{\textit{$t$}}} \left.\psi _n\otimes \psi _m^*\right)\end{equation}

\(\text{Following}\text{  }\text{time}\text{  }\text{\textit{$t$}}\text{\textit{  }}\text{the}\text{  }\text{reception}\text{  }\text{of}\text{ 
}a\text{  }\text{message}\text{  }\text{before}\text{  }\text{its}\text{  }\text{contents}\text{  }\text{or}\text{  }\text{even}\text{  }\\
\text{its}\text{  }\text{length}\text{  }\text{is}\text{  }\text{known}\text{  }\text{changes}\text{  }\text{the}\text{  }\text{state}\text{  }\text{to}\text{
 }\text{the}\text{  }\text{mixed}\text{  }\text{state}\text{  }\sum _{n=0}^{\infty } \psi _n\otimes \psi _n^* ,\text{  }\\
\text{which}\text{  }\text{is}\text{  }\text{independent}\text{  }\text{of}\text{  }\text{\textit{$t$}}.\text{  }\text{If}\text{  }\text{the}\text{
 }\text{message}\text{  }\text{length}\text{  }\text{is}\text{  }\text{known}\text{  }\text{to}\text{  }\text{be}\text{  }\text{precisely}\text{
 }\text{\textit{$k$}}\text{  }\\
\text{then}\text{  }\text{the}\text{  }\text{state}\text{  }\text{becomes}\text{  }\text{the}\text{  }\text{pure}\text{  }\text{state}\text{  } \psi
_k\otimes \psi _k^* .\text{  }\text{The}\text{  }\text{mixed}\text{  }\text{state}\text{   }\sum _{n=0}^{\infty } \psi _n\otimes \psi _n^* \\
\text{represents}\text{  }\text{the}\text{  }\text{component}\text{  }\text{of}\text{  }\text{the}\text{  }\text{state}\text{  }\text{simutaneously}\text{
 }\text{measurable}\text{  }\text{with}\text{  }\text{the}\text{  }\\
\text{message}\text{  }\text{length}.\text{  }\text{Note}\text{  }\text{that}\text{  }\text{within}\text{  }\text{this}\text{  }\text{dynamical}\text{
 }\text{model}\text{  }\text{it}\text{  }\text{is}\text{  }\text{stationary}\text{  }\text{in}\text{  }\text{time}.\)

\(\text{More}\text{    }\text{generally} \text{if}\text{    }\varrho \text{\textit{    }}\text{is}\text{    }\text{an}\text{    }\text{arbitrary}\text{
   }\text{state} \\
(i.e.,\text{    }\text{positive}\text{    }\text{operator}\text{    }\text{on}\text{    }\mathcal{H}\text{    }\text{with}\text{    }\text{finite}\text{
    }\text{positive}\text{     }\text{trace})\text{    } \text{then}\text{    }\)

\begin{equation}\text{         }\varrho \text{\textit{$ $}}\text{\textit{$=$}}\text{\textit{$ $}}\sum _{n=0}^{\infty } \text{\textit{$\varrho $}}_{\text{nn}}
+ \sum _{m\neq n} \varrho _{\text{mn}\text{    }},\text{where}\text{    }\text{\textit{$\varrho $}}_{\text{mn}}= M\left(\text{\textit{FG}}(2)_m\right)
\varrho  M\left(\text{\textit{FG}}(2)_n\right)\text{     }\end{equation}

\(\text{The}\text{  }\text{first}\text{  }\text{sum}\text{  }\text{in}\text{  }(5)\text{  }\text{represents}\text{  }\text{the}\text{  }\text{component}\text{
 }\text{of}\text{  }\varrho \text{  }\text{which}\text{  }\text{is}\text{  }\text{co-measurable}\text{  }\text{with}\text{  }\text{\textit{$N$}}.
\\
 \text{Moreover}, \text{the}\text{  }\text{expansion}\text{  }(5)\text{  }\text{is}\text{  }\text{orthogonal}\text{  }\text{with}\text{  }\text{respect}\text{
 }\text{to}\text{  }\text{the}\text{  }\text{Hilbert-Schmidt}\text{   }\text{inner}\text{  }\text{product} \\
(A,B)\rightarrow \text{Trace}(A^* B), \text{so}\)

\[  \text{Trace}\left(\varrho ^2\right.) = \sum _{n=0}^{\infty } \text{Trace}\left(\text{\textit{$\varrho $}}_{\text{nn}}^2\right) + \sum _{m\neq
n} \text{Trace}( M\left(\text{\textit{FG}}(2)_m\right) \varrho \left. M\left(\text{\textit{FG}}(2)_n\right)\varrho \right)\text{\textit{$ $}}\text{\textit{$.$}}\]

This last expression suggests defining an \textit{ index of co-measurability} with respect to \textit{ N} { }by

\begin{equation}\text{\textit{index}}\text{\textit{$($}}\text{\textit{$N$}}\text{\textit{$,$}}\varrho \text{\textit{$)$}}\text{\textit{$ $}}\text{\textit{$=$}}\text{\textit{$
$}}\frac{\sum _{n=0}^{\infty } \text{Trace}\left(\text{\textit{$\varrho $}}_{\text{nn}}^2\right) }{\text{Trace}\left(\varrho ^2\right)}\text{   
    }\end{equation}

Apparently, it is customary in quantum computation to assume that the number of bits is conserved, so that transitions from a message to a message
of greater length is impossible; such states are precisely those of index 1. This index gives a measure of the extent this assumption is correct.
In particular, there are states in QMS which allow transitions which change the message length.

\subsubsection*{ { } { } { } 2) { } { }  { }\textit{ j}th Qubit Observable:}

\(\text{What}\text{     }\text{is}\text{        }\text{the}\text{    }\text{value}\text{    }\text{of}\text{    }a\text{    }\text{particular}\text{
   }\text{bit}\text{    }(\text{or}\text{    }\text{extended}\text{    }\text{bit})?\text{    }\text{If}\text{    }\text{we} \\
\text{consider}\text{  }\text{the}\text{  }\text{\textit{$j$}}\text{th}\text{    }\text{extended}\text{    }\text{bit}\text{    }\text{of}\text{
   }a\text{    }\text{message}\text{  }\text{its}\text{    }\text{value}\text{    }\text{can}\text{    }\text{be}\text{    }0,1,2,3,\text{      
 }\\
\text{or}\text{    }-1\text{    }\text{representing}\text{    }\text{the}\text{    }\text{case}\text{    }\text{where}\text{    }\text{the}\text{
   }\text{message}\text{    }\text{length}\text{    }\text{is}\text{    }\text{not}\text{    }\text{more}\text{    }\text{than}\text{    }\text{\textit{$j$}},
\\
\text{so}\text{    }\text{the}\text{    }\text{\textit{$j$}}\text{th}\text{    }\text{bit}\text{    }\text{is}\text{    }\text{not}\text{    }\text{defined}.\text{
   }\text{Formally},\)

\begin{equation}X_j = (-1)M\left(\cup _{k\leq j}\right.\text{\textit{FG}}(2)_k) + \sum _{b=0}^3 b M\left(\left\{x\in \text{\textit{FG}}(2): \#x >
j\text{  }\text{and}\text{  }\text{\textit{$x$}}_j\right.\right.=\text{\textit{$b$}}\}) \end{equation}

\(\\
\text{Each}\text{  }\text{vector}\text{  }\text{of}\text{  }\text{the}\text{  }\text{form}\text{    }\delta _{y\text{\textit{$a$}}z} \text{  }\text{where}\text{
  }\text{\textit{$y$}},\text{\textit{z }}\in \text{\textit{FG}}\text{(2) }\\
\text{and} \text{ $\#$}\text{\textit{$y$}}=\text{\textit{$j$}} \text{   }\text{is}\text{  }\text{an}\text{  }\text{eigenvector} \text{of} \text{\textit{$X$}}_j\text{
 }\text{for}\text{  }\text{the}\text{  }\text{eigenvalue}\text{  }\text{\textit{$a$}}\in 4 ; \\
\text{the}\text{  }\text{corresponding}\text{  }\text{state}\text{  }\text{is}\text{  }\text{such}\text{  }\text{that}\text{  }\text{the}\text{ 
}\text{measurement}\text{  }\text{of}\text{  }\text{the}\text{  }\text{\textit{$j$}}\text{th}\text{  }\text{extended}\text{   }\text{bit}\text{ 
 }\text{always}\text{  }\\
\text{yields}\text{  }\text{\textit{$a$}}\text{\textit{$.$}}\text{\textit{  }}\text{Observe}\text{  }\text{that}\text{  }\text{the}\text{   }\text{operators}\text{
 }\left.\left(X_j\right.\right)_{j\geq 0} \text{together}\text{   }\text{with}\text{  } \text{\textit{$N$}}\text{\textit{  }}\text{determine}\text{
       }\text{the}\text{  }\\
\text{atomic}\text{  }\text{measure}\text{  }\text{\textit{$M$}}\text{\textit{  }}\text{completely}\text{  }\text{since}\text{  }\text{we}\text{
 }\text{know}\text{  }\text{the}\text{  }\text{message}\text{  }\text{if}\text{  }\text{we}\text{  }\text{know}\text{  }\text{the}\text{  }\text{length}\text{
 }\\
\text{and}\text{  }\text{the}\text{  }\text{value}\text{  }\text{of}\text{  }\text{the}\text{  }\text{extended}\text{  }\text{qubit}\text{  }\text{at}\text{
 }\text{every}\text{  }\text{position}\text{  }\text{less}\text{  }\text{than}\text{  }\text{the}\text{  }\text{length}.\text{   }\text{Finally},\text{
  }\\
\text{the}\text{  }\text{\textit{$j$}}\text{th}\text{  }\text{qubit}\text{  }\text{is}\text{  }\text{represented}\text{  }\text{by}\text{  }\text{the}\text{
 }\text{observable}\)

\begin{equation}\text{    }Q_j =  -M\left(\left\{x: \#x \leq  j\text{    }\text{or}\text{    }x_{j }\right.\right.\in \{2,3\}) +\sum _{b=0}^1 b M\left(\left\{x\in
\text{\textit{FG}}(2): \#x > j\text{  }\text{and}\text{  }\text{\textit{$x$}}_j\right.\right.=\text{\textit{$b$}}\})  \text{   }\end{equation}

\subsubsection*{ { } { } { } { }3) { } { } { } { } Quantum Message { }Source:}

\( \text{The}\text{    }\text{component}\text{    }\text{of}\text{    }\text{state}\text{    }\varrho \text{\textit{    }}\text{that}\text{    }\text{is}\text{
   }\text{co-measurable}\text{    }\text{with}\text{    }\text{\textit{$N$}}\text{\textit{    }}\text{is}\text{    }\text{the}\text{    }\text{state}\text{
 }\varrho _{\text{\textit{$N$}}}\text{   }.\text{     }\text{The}\\
 \text{states}\text{    }\varrho \text{\textit{    }}\text{and}\text{    }\varrho _N\text{    }\text{are}\text{    }\text{identical}\text{    }\text{as}\text{
   }\text{message-generating}\text{    }\text{sources},\text{    }\text{since}\)

\[ \text{for}\text{     }x\in \text{\textit{FG}}(2)_n , \left\langle \delta _{\text{\textit{$x$}}}|\varrho \left(\delta _{\text{\textit{$x$}}}\right)\right\rangle
 = \left\langle \delta _{\text{\textit{$x$}}}|\varrho _n\left(\delta _{\text{\textit{$x$}}}\right)\right\rangle  = \left\langle \delta _{\text{\textit{$x$}}}|\varrho
_N\left(\delta _{\text{\textit{$x$}}}\right)\right\rangle \text{    }. \]

\(\\
\text{Now}\text{  }\text{every}\text{  }\text{quantum}\text{  }\text{state}\text{  }\text{is}\text{  }\text{also}\text{  }\text{an}\text{  }\text{observable}\text{
 }\text{because}\text{  }\text{it}\text{  }\text{is}\text{  }a\text{  }\text{self-adjoint}\text{  }\text{operator} \\
\text{ $\varrho $}\text{\textit{ = }}\sum _r \text{\textit{$r$}} E_r \text{    } \text{where}\text{    }\text{the}\text{    }\text{orthogonal}\text{
   }\text{sum}\text{    }\text{is}\text{    }\text{over}\text{    }\text{\textit{$r$}}\text{ $\in $ spectrum($\varrho $)}\text{\textit{$.$}}\text{\textit{
    }}\text{The}\text{    }\\
\text{possible}\text{    }\text{measured}\text{    }\text{values}\text{    }\text{are}\text{    }\text{the}\text{    }\text{elements}\text{     }\text{\textit{$r$}}.\text{
   }\text{If}\text{    }\text{we}\text{    }\text{measure}\text{    }\varrho \text{    }\text{while}\text{    }\text{in}\text{    }\text{the}\text{
   }\\
\text{state}\text{    }\varrho \text{\textit{        }}\text{the}\text{    }\text{probability}\text{    }\text{that}\text{    }\text{the}\text{ 
  }\text{outcome}\text{    }\text{is}\text{      }\text{\textit{$r$}}\text{       }\text{equals}\text{  } m_{r\text{  }}r/\text{Trace}(\varrho ),\text{
   }\\
\text{where}\text{  }\text{\textit{$m$}}_r\text{  }\text{is}\text{  }\text{the}\text{  }\text{multiplicity}\text{  }\text{  }\text{of}\text{    }\text{\textit{$r$}}\text{
   }\text{as}\text{    }\text{an}\text{    }\text{eigenvalue}\text{    }\text{of}\text{    }\varrho \text{\textit{$.$}}\text{\textit{    }}\text{The}\text{
   }\text{\textit{$a$}}\text{\textit{    }}\text{\textit{posteriori}}\text{\textit{    }}\text{state}\text{    }\\
\text{is}\text{   }r E_r\text{    }\text{and}\text{    }\text{further}\text{    }\text{measurements}\text{    }\text{of}\text{    }\varrho \text{\textit{
   }}\text{always}\text{    }\text{yield}\text{        }\text{\textit{$r$}}\text{        }\text{with}\text{    }\text{no}\text{    }\text{change}\text{
   }\\
\text{of}\text{    }\text{state}.\text{  }\text{In}\text{    }\text{the}\text{    }\text{special}\text{    }\text{case}\text{    }\text{that}\text{
   }\varrho \text{\textit{    }}\text{is}\text{    }a\text{    }\text{pure}\text{    }\text{state}\text{    }\psi \otimes \psi ^*\text{    }\text{the}\text{
   }\text{measurement}\text{    }\\
\text{always}\text{   }\text{yields}\text{    }\left|\psi |^2\right.\text{    }\text{with}\text{    }\text{probability}\text{    }\text{one}.\text{
   }\text{As}\text{    }\text{we}\text{    }\text{have}\text{    }\text{seen}\text{    }\text{in}\text{  }\text{this}\text{  }\text{case}\text{ 
  }\varrho _{N\text{    }}=\text{    }\\
\sum _{n=0}^{\infty } \psi _n\otimes \psi _n^*,\text{    }\text{and}\text{    }\text{this}\text{    }\text{is}\text{    }\text{always}\text{    }a\text{
   }\text{mixed}\text{    }\text{state}\text{    }\text{unless}\text{    }\psi \text{    }\text{is}\text{    }\text{an}\text{  }\text{eigenvector}\text{
   }\\
\text{of}\text{    }\text{\textit{$N$}}\text{\textit{$.$}}\text{\textit{    }}\text{Now}\text{    }\text{if}\text{    }\text{we}\text{    }\text{measure}\text{
   }\psi \otimes \psi ^*\text{    }\text{the}\text{    }\text{result}\text{    }\text{is}\text{    }\left|\psi |^2\right.\text{    }\text{with}\text{
   }\text{probability}\)

\[\text{    }\text{\textit{index}}\text{\textit{$($}}\text{\textit{$N$}}\text{\textit{$,$}}\psi \otimes \psi ^*)\text{    }=\text{    }\sum _{n=0}^{\infty
} \left|\psi _n|^4\right./\left|\psi |^4\right. < 1\text{    }\text{in}\text{    }\text{general}.\text{\textit{         }}\]

\subsubsection*{ { } { } { } { } { } 4) { } { } Some { }Examples;}

\(.\text{\textit{$ $}}\text{According}\text{    }\text{to}\text{    }\text{results}\text{    }\text{in}\text{    }[\text{KB}]\text{     }\text{the}\text{
   }\text{maximum}\text{    }\text{entropy}\text{    }\text{probability}\text{    }\\
\text{measure}\text{    }\text{on}\text{    }\text{\textit{FG}}(2)\text{    }\text{with}\text{    }a\text{    }\text{fixed}\text{    }\text{average}\text{
       }\text{message}\text{    }\text{length}\text{    }\mu \text{            }\text{is}\text{    }\\
\text{given}\text{    }\text{by} \text{ w(x) = }(1-p)\left.p^n\right/\#\text{\textit{FG}}(2)_n\text{  }\text{where}\text{   }\text{n = $\#$x} \text{
    }\text{and}\text{    } \text{\textit{$p$}} = \frac{\mu }{1+\mu }.\text{        }\\
\text{Let}\text{    } \psi  = \sum _{\text{x$\epsilon $FG}(2)} \sqrt{\text{\textit{$w$}}(\text{\textit{$x$}})} \delta _x.\text{    }\text{As}\text{
   }a\text{    }\text{message}\text{    }\text{source}\text{    }\text{the}\text{        }\text{pure}\text{    }\\
\text{state}\text{    }\psi \otimes \psi ^*\text{    }\text{is}\text{    }\text{such}\text{    }\text{that}\text{    }\text{the}\text{    }\text{probability}\text{
   }\text{of}\text{    }\text{receiving}\text{    }\text{\textit{$x$}}\text{    }\text{is}\text{    }\text{\textit{$w$}}(\text{\textit{$x$}}).\text{
  }\)

\( \text{Now}\text{    }\text{consider}\text{    }\text{the}\text{    }\text{highly}\text{    }\text{mixed}\text{    }\text{state}\text{    }\text{
   }\sigma =\sum _{x } w(x)\delta _x\otimes \delta _x^*.\text{    }\text{As}\text{    }\\
a\text{    }\text{message}\text{    }\text{source}\text{    }\sigma \text{    }\text{induces}\text{    }\text{the}\text{    }\text{same}\text{  
 }\text{probability}\text{    }\text{distribution}\text{    }\\
\text{on}\text{    }\text{the}\text{    }\text{messages}\text{    }\text{as}\text{        }\psi \otimes \psi ^* . \text{The}\text{    }\text{entropy}\text{
   }\text{of}\text{    }\text{this}\text{    }\text{distribution}\text{    }\text{is}\text{    }\\
\text{log(3/p) p/(1-p) + log(4/3) - log(1-p)}; \text{  }\text{as}\text{    }\mu \to \infty \text{    }\text{the}\text{    }\text{entropy}\text{ 
  }\text{is}\text{    }\text{asymptotic}\text{    }\text{to}\text{    }\\
\log (3)\mu \text{    }.\text{    }\)

\( \text{But}\text{  }\text{the}\text{    }\text{von}\text{    }\text{Neumann}\text{  }\text{entropy}\text{    }\text{for}\text{     }\psi \otimes
\psi ^*\text{  }\text{is}\text{       }\text{zero}\text{    }\text{because}\text{    }\text{it}\text{  }\text{is}\text{  }a\text{    }\text{pure}\text{
   }\text{state},\text{  }\\
\text{whereas}\text{    }\text{the}\text{    }\text{von}\text{    }\text{Neumann}\text{    }\text{entropy}\text{    }\text{of}\text{    }\sigma \text{
   }\text{equals}\text{    }\text{the}\text{    }\text{source}\text{    }\text{entropy}.\)

\subsection*{\\
D. Positive Messages}

\(\text{Let}\text{  }\mathcal{H}^+\text{  }\text{be}\text{  }\text{the}\text{  }\text{Hilbert}\text{  }\text{space}\text{  }\text{spanned}\text{
 }\text{by}\text{   }\{\delta _y: y \epsilon  2^*\}.\text{  }\mathcal{H}^+  \text{is}\text{  }\\
\text{the}\text{  }\text{positive}\text{  }\text{message}\text{  }\text{state}\text{  }\text{space}\text{  }\text{of}\text{  }\text{ordinary}\text{
 }\text{bit-string}\text{  }\text{messages}.\text{  }\text{The}\text{  }\\
\text{orthogonal}\text{  }\text{projector}\text{  }\text{onto}\text{  }\mathcal{H}^+\text{  }\text{is}\text{  }M\left(2^*\right)\text{as in (1) above}.
\text{The}\text{  }\text{operator} \\
\text{\textit{$N$}}^+\text{ = }M\left(2^*\right)\text{\textit{$N$}}\text{\textit{$ $}}M\left(2^*\right) \text{has}\text{  }\text{many}\text{  }\text{properties}\text{
 }\text{analogous}\text{  }\text{to}\text{  }\text{those}\text{  }\text{of}\text{  }\text{\textit{$N$}}. \\
\text{The}\text{  }\text{state}\text{  }\text{change}\text{  }\text{wrought}\text{  }\text{by}\text{  }a\text{  }\text{test}\text{  }\text{for}\text{
 }\text{positivity}\text{  }\text{with}\text{  }\text{no}\text{  }\text{information}\text{  }\text{about}\text{  }\text{the}\text{  }\text{outcome}
\\
\text{changes}\text{  }\varrho \text{  }\text{to}\text{  }M\left(2^*\right)\text{$\varrho $ }M\left(2^*\right)+\left(\text{\textit{$I$}}-\left.M\left(2^*\right)\right)\text{$\varrho
$(}\text{\textit{$I$}}- \left.M\left(2^*\right)\right)\right.\text{and}\text{  }\text{the}\text{  }\text{trace}\text{  }\text{is}\text{  }\text{preserved},
\\
\text{but}\text{  }a\text{  }\text{state}\text{  }\text{change}\text{  }\text{from}\text{  }\varrho \text{   }\text{to}\text{   }M\left(2^*\right)\left.\varrho
\text{\textit{$M$}}(2^*\right)\text{which}\text{  }\text{occurs}\text{   }\text{when}\text{  }\text{the}\text{  }\text{test}\text{  }\text{result}\text{
 }\\
\text{is}\text{  }\text{affirmative}\text{  }\text{will}\text{  }\text{of}\text{  }\text{course}\text{  }\text{provide}\text{  }\text{information}\text{
 }\text{and}\text{  }\text{reduce}\text{  }\text{the}\text{  }\text{trace}\text{  }\text{of}\text{  }\text{the}\text{  }\text{state}.\)

\(\text{Multiplication}\text{  }\text{in} \text{\textit{FG}}\text{(2)} \text{  }\text{restricted}\text{  }\text{to}\text{   }2^*\text{  }\text{is}\text{
 }\text{merely}\text{  }\text{concatenation}\text{   }\text{of}\text{   }\text{bit}\text{  }\text{strings}.\text{   }\text{The}\\
\text{  }\text{multiplication}\text{  }\text{derived}\text{  }\text{from}\text{  }\text{the}\text{  }\text{concatenation}\text{  }\text{product}\text{
 }\text{of}\text{  }\text{basis}\text{  }\text{vectors}\text{  }\text{indexed}\text{  }\text{by}\text{  }\\
\text{multi-qubits}\text{  }\text{extends}\text{  }\text{to}\text{  }a\text{  }\text{product}\text{  }\text{on}\text{   }\mathcal{H}^+\text{  }\text{which}\text{
 }\text{is}\text{  }\text{the}\text{  }\text{completion} \text{of}\text{  }\text{the}\text{  }\text{tensor}\text{   }\text{algebra}\text{  }\text{on}\text{
 }\text{the}\\
 \text{two-dimensional}\text{  }\text{Hilbert}\text{  }\text{space}\text{  }\text{spanned}\text{  }\text{by}\text{  }\text{the}\text{  }\text{reference}\text{
 }\text{qubits} |0\rangle \text{  }\text{and} |1\rangle .\text{   }\text{In}\text{   }\text{quantum}\text{  }\\
\text{computation} \text{\textit{$$}}   \delta _y \text{  }\text{is}\text{  }\text{often}\text{  }w\text{ritten}\text{   }\text{as}\text{   }|\text{\textit{$y$}}\rangle
\text{   }\text{\textit{$ $}}(\text{ket}\text{   }\text{form})\text{   }\text{and}\text{  } \delta _y^*\text{   }\text{as}\text{   }\langle \text{\textit{$y$}}|\text{
 }(\text{bra}\text{   }\text{form})\text{    }\text{so}\text{   }\text{for}\text{   }\text{example}\\
\text{  }\delta _{1101}\otimes \delta _{1101}^*  \text{may}\text{   }\text{be}\text{  }\text{written}\text{  }\text{more}\text{  }\text{compactly}\text{
 }\text{as}\text{  }|\text{\textit{$1101$}}\rangle \langle \text{\textit{$1101$}}|\text{  } . \)

\(\text{  }\text{Here}\text{  }\text{is}\text{  }a\text{  }\text{sample}\text{  }\text{computation}:\)

\[( -1 |0\rangle  + 3 |00\rangle  - i|10\rangle  )*( |\Lambda \rangle  +\text{  }|1\rangle  + |01\rangle  ) =\text{  }- |\text{\textit{$0$}}\text{\textit{$\rangle
$}} + 3 |\text{\textit{$00$}}\text{\textit{$\rangle $}}\text{\textit{$ $}} - i\text{\textit{$ $}}|\text{\textit{$10$}}\text{\textit{$\rangle $}}\text{\textit{$
$}} - |\text{\textit{$01$}}\text{\textit{$\rangle $}}\text{\textit{$ $}} + 2 |\text{\textit{$001$}}\text{\textit{$\rangle $}}\text{  }-i\text{\textit{$
$}}|\text{\textit{$101$}}\text{\textit{$\rangle $}}\text{  }+ 3 |\text{\textit{$0001$}}\text{\textit{$\rangle $}}\text{  }- i\text{\textit{$|$}}\text{\textit{$1001$}}\text{\textit{$\rangle
$}}\]

\(\text{This}\text{   }\text{multiplication}\text{  }\text{is}\text{  }\text{precisely}\text{   }\text{convolution}\text{  }\text{on}\text{  }\text{\textit{FG}}(2).\text{
 }\text{The}\text{   }\text{QMS}\text{    }\mathcal{H}\text{   }\text{is}\text{  }\text{the}\text{  }\text{extension}\text{  }\\
\text{of}\text{  }\mathcal{H}^+ \text{  }\text{analogous}\text{  }\text{to}\text{  }\text{the}\text{  }\text{extension}\text{  }\text{of}\text{ 
}\text{the}\text{  }\text{semigroup}\text{  }2^*\text{  }\text{to}\text{  }\text{the}\text{  }\text{group}\text{  }\text{\textit{FG}}(2). \)

On the other hand, perhaps some physical constraint restricts our message measurements to \(2^*\). { }Many { } data-processing operations \textit{
F} in QMS would have as a goal the maximization of { }Trace(M(\(2^*\))\textit{ F}($\varrho $)M(\(2^*\)) )/Trace($\varrho $)  subject to some bounds
on Trace($\varrho $). However, the operator \textit{ F} need not preserve states supported { }by { }\(\mathcal{H}^+\) .  { }The positive message
observable's spectral measure is \(M^+\) , which is the atomic measure for the observable taking values in \(2^*\)$\cup $$\{$(-)$\}$, where (-) {
}stands for any fixed thing not in \(2^*\) :

\[ M^+(\{a\}) = \delta _a\otimes  \delta _a^*\text{  }\text{for}\text{  }a\in 2^*\text{  },  M^+(\{(-)\}) = \left.M\left(\text{\textit{FG}}(2)\sim
2^*\right.\right)=I-M^+\left(2^*\right.) .\]

\(\text{Quantum}\text{   }\text{message}\text{  }\text{space}\text{  }\text{contains}\text{  }\text{all}\text{  }\text{the}\text{  }\text{usual}\text{
 }\text{spaces}\text{   }\text{for}\text{  }\text{quantum}\text{  }\text{circuits}: \)

\[\text{Span}\left\{\delta _x\right.:x\in 2^n\} \left.\left.=\text{Range}\left(M\left(2^n\right.\right.\right)\right)\approx \mathbb{C}^{2^n}\approx
 \mathbb{C}^{\otimes n},\text{  }\text{the}\text{  }\text{space}\text{  }\text{of}\text{  }\text{\textit{n-}}\text{qubits}. \]

QMS { }can absorb the tensor product of two such spaces such as  \(\mathbb{C}^{\otimes k}\) representing \textit{ k} control bits with \(\mathbb{C}^{\otimes
m}\) data bits. The usual quantum operators of conventional quantum circuit theory can all be extended trivially to \(\mathcal{H}^+\) ; the extensions
to $\mathcal{H}$ are not so obvious.

\subsection*{E. Data Processing on Messages}

\(A\text{   }\text{data-processing}\text{  }\text{operation}\text{  }\text{on}\text{  }\text{states}\text{  }\text{in}\text{  }\text{QMS}\text{ 
}\text{is}\text{  }\text{implemented}\text{  }\text{by}\text{  }a\text{  }\text{\textit{quantum}}\text{\textit{$ $}}\text{\textit{operation}}\text{\textit{$
$}}[\text{NC}],\text{  }\\
\text{which}\text{  }\text{in}\text{  }\text{this}\text{  }\text{context}\text{  }\text{means}\text{   }a\text{  }\text{real}\text{  }\text{linear}\text{
 }\text{operator}\text{  }\text{\textit{$F$}}\text{\textit{  }}\text{on}\text{  }\text{the}\text{  }\text{self-adjoint}\text{  }\text{operators}\text{
 }\text{of}\text{  }\\
\text{trace}\text{  }\text{class}\text{  }\text{on}\text{  }\mathcal{H}\text{  }\text{which}\text{  }\text{is}\text{  }\text{strictly}\text{  }\text{positive}\text{
 }\text{and}\text{  }\text{non-trace-increasing}.\text{  }\text{In}\text{  }\text{particular},\text{  }\\
\text{if}\text{  }\varrho \text{  }\text{is}\text{  }a\text{  }\text{message}\text{  }\text{state}\text{  }\text{then}\text{  }\text{so}\text{  }\text{is}\text{
 }\text{\textit{$F$}}(\varrho ),\text{  }\text{and}\text{  }\text{0 $<$ Trace(}\text{\textit{$F$}}\text{($\varrho $)) $\leq $ Trace($\varrho $)}.\text{
 }\text{The}\text{   }\\
\text{information}\text{  }\text{lost}\text{  }\text{by}\text{  }\text{the}\text{  }\text{system} (\text{or}\text{  }\text{entropy}\text{  }\text{gained}?)\text{
 }\text{is}\text{  }-\log _2\left(\frac{\text{Trace}(F\text{\textit{$ $}}(\varrho ))}{\text{Trace}(\varrho )}\right).\text{  }\\
\text{If}\text{  }\text{equality}\text{  }\text{holds}\text{  }\text{for}\text{  }\text{all}\text{  }\text{states}\text{   }\varrho \text{   }\text{then}\text{
 }\text{\textit{$F$}}\text{  }\text{is}\text{  }\text{trace-preserving};\text{  }\\
\text{the}\text{  }\text{interpretation}\text{  }\text{is}\text{  }\text{that}\text{  }\text{the}\text{  }\text{state}\text{  }\text{change}\text{
 }\varrho \to \text{\textit{$F$}}\text{($\varrho $)}\text{   }\text{generates}\text{  }\text{no}\text{  }\text{information},\text{  }\\
\text{does}\text{  }\text{not}\text{  }\text{increase}\text{  }\text{entropy},\text{  }\text{and}\text{  }\text{is}\text{  }\text{often}\text{  }\text{reversible}.\text{
 }\text{An}\text{  }\text{example}\text{  }\text{of}\text{  }\text{such}\text{  }\text{an}\text{  }\text{operation}\text{  }\text{is}\text{  }\\
\mathcal{K}(U)(\varrho ) = \text{\textit{$U$}}\varrho \text{\textit{$U$}}^*,\text{  }\text{where}\text{  }U\text{  }\text{is}\text{  }a\text{   }
\text{unitary}\text{  }\text{operator}\text{  }\text{on}\text{  }\text{the}\text{  }\text{QMS}\text{   }\mathcal{H}.\text{   }\text{An}\text{  }\text{example}\text{
 }\\
\text{of}\text{  }a\text{  }\text{quantum}\text{  }\text{operation}\text{  }\text{which}\text{  }\text{is}\text{  }\text{not}\text{  }\text{reversible}\text{
 }\text{is}\text{  }\text{the}\text{  }\text{measurement}\text{  }\text{of}\text{  }\text{an}\text{  }\text{observable},\text{  }\\
\text{of}\text{  }a\text{  }\text{projector}\text{  }\text{\textit{$P$}}\text{  }\text{say},\text{  }\text{with}\text{  }1\text{  }\text{as}\text{
 }\text{the}\text{  }\text{result}\text{  }\text{of}\text{  }\text{the}\text{  }\text{measurement}.\text{  }\text{We}\text{  }\text{have}\text{ $\mathcal{Q}$(}\text{\textit{$P$}}\text{)($\varrho
$)=}\text{\textit{$P$}}\varrho \text{\textit{$P$}} .\text{  }\text{If}\text{  }\\
\text{we}\text{  }\text{measure}\text{  }\text{\textit{$P$}}\text{\textit{  }} \text{without}\text{  }\text{changing}\text{  }\text{the}\text{  }\text{entropy}\text{
 }\text{of}\text{  }\text{the}\text{  }\text{system}\text{  }t \text{he}\text{  }\text{\textit{$a$}}\text{\textit{  }}\text{\textit{posteriori}}\text{\textit{
 }}\text{state}\text{  }i s\)

\[\mathcal{T}(\text{\textit{$P$}})(\varrho )=\mathcal{Q}(\text{\textit{$P$}})(\varrho )+\mathcal{Q}(\text{\textit{$I$}}-\text{\textit{$P$}})(\varrho
)\]

\(\text{Certain} o\text{ther}\text{   }\text{trace-preserving}\text{   }\text{operations}\text{   }\text{can}\text{   }\text{be}\text{   }\text{built}\text{
  }\text{up}\text{   }\text{from}\text{   }\text{orthogonal}\text{   }\text{projectors}  \left(P_j\right.) \\
\text{such}\text{   }\text{that}\text{  } \sum _j P_j=\text{\textit{$I$}}\text{\textit{   }}\text{and}\text{   }\text{associated}\text{   }\text{unitary}\text{
  }\text{operators}\text{   }\left(U_j\right.):\)

\[\text{  }\text{let}\text{  }\mathcal{T}(U,P)(\varrho ) = \sum _j U_jP_j\text{$\varrho $P}_jU_j^*, \text{assuming}\text{   }\text{the}\text{   }\text{range}\text{
  }\text{of}\text{   }\text{the}\text{   }\text{index}\text{   }\text{\textit{$j$}}\text{\textit{  }} \text{is}\text{   }\text{finite}.\]

\( \text{If}\text{   }a\text{   }\text{measurement}\text{    }\text{of}\text{   }\text{the}\text{   }\text{index}\text{   }\text{is}\text{   }\text{made}\text{
  }\text{with}\text{   }\text{result}\text{  }\text{\textit{$k$}}\text{\textit{  }}\text{the}\text{  }\text{state}\text{  }\text{becomes}\text{ 
}P_k\varrho \text{\textit{$P$}}_k\text{  }\text{and}\text{  }\\
 \text{\textit{$\mathcal{T}$}}\text{\textit{$(U,P)$}}\text{\textit{$ $}}( P_k\varrho \text{\textit{$P$}}_k) = U_kP_k\varrho \text{\textit{$P$}}_kU_k^*,\text{
  }\text{so}\text{   }\text{after}\text{   }\text{measuring}\text{   }\text{and}\text{  }\text{knowing}\text{   }\\
\text{the}\text{   }\text{result}\text{   }\text{the}\text{   }\text{operator}\text{   }\text{\textit{$\mathcal{T}$}}\text{    }\text{does}\text{
  }a\text{   }\text{switch  } \text{operation},\text{   }\text{applying}\text{   }\mathcal{K}\left(U_k\right.)\text{   }\\
\text{to}\text{   }\text{the}\text{   }\text{\textit{$a$}}\text{\textit{   }}\text{\textit{posteriori}}\text{\textit{   }}\text{state}\text{   }\text{conditionally}.\text{
   }\text{However}\text{  }\mathcal{T}\text{\textit{$(U,P)$}}\text{   }\text{may}\text{   }\text{be}\text{   }\text{applied}\text{   }\\
\text{without}\text{   }\text{ever}\text{   }\text{actually}\text{   }\text{getting}\text{   }\text{the}\text{  }\text{result}\text{   }\text{of}\text{
  }\text{the}\text{   }\text{measurement}. \)

\(\text{In}\text{   }\text{the}\text{   }\text{special}\text{   }\text{case}\text{   }\text{when} \text{  }U_j\text{   }\text{and}\text{  } P_j\text{
  }\text{commute}\text{   }\text{for}\text{   }\text{all}\text{   }\text{\textit{$j$}}\text{\textit{    }}\text{then}\text{   }\\
\text{ $\mathcal{T}$(}\text{\textit{$U$}},\text{\textit{$P$}}\text{) = $\mathcal{K}$(}\text{\textit{$W$}}\text{)$\circ \mathcal{T}$(}\text{\textit{$I$}},\text{\textit{$P$}}\text{)
= $\mathcal{T}$(}\text{\textit{$I$}},\text{\textit{$P$}}\text{)$\circ \mathcal{K}$(}\text{\textit{$W$}})\text{  }\text{where}\text{   }\mathcal{T}(\text{\textit{$I$}},\text{\textit{$P$}})=\sum
_{\text{\textit{$j$}}} \mathcal{Q}\left(P_j\right)\text{   }\text{is}\text{  }\\
a\text{  }\text{measurement}\text{   }\text{of}\text{   }\text{the}\text{   }\text{outcome}\text{   }\text{\textit{$j$}}\text{   }\text{without}\text{
  }\text{any}\text{   }\text{definite}\text{    }\text{information}\text{   }\text{about}\text{   }\text{its} \text{value}.\)

\subsection*{F. { }Definite Message States as Registers\\
}

\subsubsection*{ { } { }1) The Destructive Read Memory Cell: }

\textit{ \\
} The cell is in the state \(\delta _b\) where \textit{ b $\in $ 2} . We need to read the bit stored, \textit{ b, }and restore it by quantum operations.
The cell requires the measurement of whether there is a message or not (i.e., the observable \(\delta _{\Lambda }\)$\otimes $\(\delta _{\Lambda }^*\))
. The unitary operators necessary are the ones which append the bits and which erase them :\\
 { }\(V_0\),\(V_1\),\(V_2\),\(V_{3 }\) , where \(V_z\) { }is the unique unitary operator which maps \(\delta _x\) to \(\delta _{x\cdot z}\).

To read the value of \textit{ b } we may proceed:\\
\hspace*{5.5ex} 1.a.1) Apply \(V_3\) to { }\(\delta _b\) to get { }\(V_3\)(\(\delta _b\)), which equals \(\delta _{\Lambda }\) if \textit{ b}=1 but
equals \(\delta _{03}\) if \textit{ b}=0 . { } { } { } { } { } \\
\hspace*{5.5ex} 1.a.2) Determine if the bit was erased by measuring \(\delta _{\Lambda }\)$\otimes $\(\delta _{\Lambda }^*\). If the result is 1
then a bit was erased which must have had the value 1; { }if the result is 0 the state is \(\delta _{03}\) so \textit{ b }must be 0. In either case
the result of measuring { }\(\delta _{\Lambda }\)$\otimes $\(\delta _{\Lambda }^*\) { }is the value of \textit{ b. { } { } { } { } { } { } { } {
} { } { } { } { }\\
 { } { } { } { } { } }1.a.3) Knowing \textit{ b, }apply \(V_b\) to restore the original value of the cell.\\
\hspace*{5.5ex} The quantum operation for reading the bit is $\mathcal{K}$(\(V_b\))$\circ \mathcal{Q}$(\(\delta _{\Lambda }\)$\otimes $\(\delta _{\Lambda
}^*\))$\circ $$\mathcal{K}$(\(V_3\)). { } { } { } { } { } { } { } { } { } 

To write the value \textit{ a}$\in $2 to the cell, first erase as before:\\
\hspace*{5.5ex} 1.b.1) Apply \(V_3\) to { }\(\delta _b\) to get { }\(V_3\)(\(\delta _b\)), which equals \(\delta _{\Lambda }\) if \textit{ b}=1 but
equals \(\delta _{03}\) if \textit{ b}=0 . { } { } { } { } { } { } { }\\
\hspace*{5.5ex} 1.b.2) Determine if the bit was erased by measuring \(\delta _{\Lambda }\)$\otimes $\(\delta _{\Lambda }^*\); the result is \\
\hspace*{5.5ex} the value of \textit{ b.}\\
\hspace*{5.5ex} 1.b.3) { } If \textit{ b}=1 Apply \(V_a\) to \(\delta _{\Lambda }\)to get \(\delta _a\)\\
\hspace*{6.ex} 1.b.4) If \textit{ b}=0 Apply \(V_a\)$\circ $\(V_2\)$\circ $\(V_1\) to \(\delta _{03}\)to get \(\delta _a\)\\
The quantum operation for writing the bit is\\
 $\mathcal{K}$(\(V_a\))$\circ $$\mathcal{T}$((\(V_2\)$\circ $\(V_1\),I),(I-\(\delta _{\Lambda }\)$\otimes $\(\delta _{\Lambda }^*\),\(\delta _{\Lambda
}\)$\otimes $\(\delta _{\Lambda }^*\)))$\circ $$\mathcal{K}$(\(V_3\)).\\
\hspace*{0.5ex} Note that in this formulation the value of \textit{ b }does not need to be known.\\

\subsubsection*{ { } { }2) { }The Shift Register}

\textit{ \\
} QMS states can store bit strings of arbitrary length, and the data may be processed rather like shift registers. A message state \(\delta _x\)
stores a bit string \(x_0\)\(x_1\)...\(x_{n-1}\)$\in $\(2^n\). We can shift bits in and out on the right by the operators $\mathcal{K}$(\(V_a\)),
\textit{ a}$\in $4. { }We also assume we can measure \textit{ N}, the message length.\\
To shift in a bit \textit{ b, }$\mathcal{K}$(\(V_b\))(\(\delta _x\)$\otimes $\(\delta _x^*\)) =\textit{ \(\text{\textit{$\delta _{\text{xb}}$}}\)}$\otimes
$\(\text{\textit{$\delta _{\text{xb}}$}}^*\). The length is now \textit{ n+1 }.\\
To shift out a bit, suppose { }\textit{ x=ab, { } }\textit{ a}$\in $\(2^{n-1}\):\\
\hspace*{1.5ex} 2.b.1) Measure \textit{ N, }and record the result \textit{ n. }This measurement doesn't change the state since we're assuming it
is an eigenstate for \textit{  N. \\
 { } }2.b.2) Erase a 1 by applying \(V_3\) to get \(\delta _{\text{\textit{$\text{ab}\cdot 3$}}}\); the state vector is either \(\delta _{\text{\textit{a03}}}\)
or \(\delta _a\).\\
\hspace*{1.5ex} 2.b.3) Now measure \textit{ N { }}again, this time getting the result \textit{ n' . }If \textit{ n'}$\leq $\textit{ n }leave the
state \(\delta _a\) alone (we know the bit shifted out was a 1); { } otherwise apply \(V_2\)$\circ $\(V_1\)to \(\delta _{\text{\textit{a03}}}\).\textit{
 }In either case the state vector is now \(\delta _a\). \\
\hspace*{6.ex} The quantum operator for shifting out the bit is { } { } { } { } { }

\( \mathcal{K}((\text{\textit{$n'\leq n$}})I + \left.\left.\left.\left.(n'>n)V_2\circ V_1\right)\circ Q\left(M\left(2^{n'}\right.\right.\right)\right)\circ
\mathcal{K}\left(V_3\right.\right)\circ \mathcal{Q}\left(M\left(2^n\right.\right.)).\\
 \)

\subsection*{G. Communication in Quantum Message Space}

 The classical Shannon model for communication over a channel is summarized by the familiar flow diagram: { } 

 { } { } { } { } { } \textit{  message { } { } { }sent signal { } { } { } { } { } { } { } { } { } { } { } { } { } { } { } { } { } { } received signal
{ } { } { } { } { } message}\\
 { } { } { } { } { }\pmb{ source $\to $ encoder/transmitter $\to $ channel $\to $ decoder/receiver $\to $ destination\\
\hspace*{36.ex} $\uparrow $\\
 { } { } { } { } { } { } { } { } { } { } { } { } { } { } { } { } { } { } { } { } { } { } { } { } { } { } { } { } { } { } { } { } { } { } }\textit{
noise { } { } { } { } { } { } { } { } { } { } { } { } { } { } { } { } { } { } { } { } { } { } { } { } { } { } { } { } { } { } { }\\
 { } { } { } { } { } { } { } { } { } { } { } { } { } { } { } { } { } { } { } { } { } { } { } { } { } { } { } { } { } { } { } { } { } { } { } { }
{ } { } { } { } { } { } { } { } { } { } { } { } { } { } { } { } { } { } { } { } { } { } { } { } { } { } { } { } { } { } { } { } }

Rather than over-stretch the analogy with the classical situation, let us describe the steps involved in Alice sending a free-group string, possibly
a binary string, to Bob:\textit{  { } { } { } { } { } { } { } { } { } { } { } { } { } { } { } { } { } { } { } { } { } { } { }}

\subsubsection*{\(\text{   }0)\text{  }\text{Start}\text{  }\text{at}\text{  }\text{the}\text{  }\text{source}. \)}

\(\text{Let}\text{   }\rho \text{   }\text{be}\text{  }\text{any}\text{  }\text{state}\text{  }\text{in}\text{  }\text{message}\text{  }\text{space}.\text{
  }\text{The}\text{  }\text{sender}\text{  }\text{Alice}\text{  }\text{gets}\text{  }\text{the}\text{  }\text{message}\text{  }\text{from}\text{
 }\text{the}\text{  }\text{source}\text{  }\text{by}\text{  }\\
\text{measuring}\text{  }\text{the}\text{  }\text{message}\text{  }\text{observable}.\text{   }\text{The}\text{  }\text{probability}\text{  }\text{that}\text{
 }\text{the}\text{  }\text{selected}\text{  }\text{message}\text{  }\text{is}\text{  }\text{\textit{$z$}}\text{\textit{  }}\text{equals}\text{  }\\
\frac{\left. \left\langle \delta _z\right.|\rho \left(\delta _z\right)\right\rangle }{\text{Tr}(\rho )}.\text{   }\text{Note}\text{  }\text{that}\text{
 }\text{the}\text{  }\text{source}\text{  }\text{may}\text{  }\text{be}\text{  }a\text{  }\text{message}\text{  }\text{that}\text{  }\text{Alice}\text{
 }\text{composed}\text{  }\text{herself}\text{  }\\
\text{by}\text{  }\text{applying}\text{  }\text{the}\text{  }\text{unitary}\text{  }\text{operators}\text{  }V_0\text{\textit{  }}\text{and}\text{
 }V_1\text{  }\text{in}\text{  }\text{proper}\text{  }\text{order}\text{  }\text{starting}\text{  }\text{with}\text{   }\delta _{\Lambda } .\text{
  }\text{For}\text{  }\\
\text{example}\text{   }\text{if}\text{  }\text{z $\in $ }\text{\textit{FG}}\text{(2)}\text{  }\text{equals}\text{  }10010\text{  }\text{then}\text{
 }\text{the}\text{  }\text{composed}\text{  }\text{source}\text{  }\text{would}\text{  }\text{be}\text{  }\text{the}\text{  }\text{pure}\text{  }\\
\text{message}\text{  }\text{state}\text{  }\text{determined}\text{  }\text{by}\text{  }\text{the}\text{  }\text{basis}\text{  }\text{vector}\text{
 } V_0 \circ \text{  }V_1 \circ  V_0 \circ \text{  }V_0 \circ  V_1\left(\delta _{\Lambda }\right.\text{\textit{$)$}}\text{\textit{  }}\text{\textit{$=$}}\text{\textit{$
$}}\delta _z .\)

\subsubsection*{\textit{ \textup{  { } { } 1) Alice gets the message to send}\textup{ . }}}

\(\text{After}\text{  }\text{this}\text{  }\text{action}\text{  }\text{the}\text{  }\text{state}\text{  }\text{is}\text{   }\left.\left.\left\langle
\delta _z\right.|\rho \left(\delta _z\right.\right)\right\rangle \delta _z\otimes \delta _z^*\text{  }\text{and}\text{  }\text{Alice}\text{  }\text{has}\text{
 }\text{the}\text{  }\text{message}\text{   }\text{\textit{$z$}}\text{   }\text{to}\text{  }\text{send}\text{  }\text{to}\text{  }\text{Bob}.\\
\text{The} \text{prabability} \text{of} \text{getting} \text{this} \text{message} \text{from} \text{source} \rho  \text{is}\)

\[\Pr (z|\rho )=\left.\left.\left.\left\langle \delta _z\right.|\rho \left(\delta _z\right.\right)\right\rangle \right/\text{Tracc}(\rho )\]

\subsubsection*{\textrm{  { } { } 2) Alice encodes the message.}}

By applying a unitary operator \textit{ G} the state is transformed so that its normalized form is { }$\psi $$\otimes $\(\psi ^*\) ; { }$\psi $ =
G(\(\delta _z\))  need not be a message basis vector nor even be a multiple qubit (i.e., { }$\psi $ need not be an eigenvector of \textit{ N ).}

\subsubsection*{\textrm{  { } { } 3) Alice transmits the encoded message over the channel. }}

The general channel is modelled as a trace-preserving quantum operator $\mathcal{E}$ mapping states on $\mathcal{H}$ to states. $\mathcal{E}$ may
have an operator sum decomposition 

\begin{equation}\mathcal{E}(\tau ) = \sum _j O(j) \tau  O(j)^*\end{equation}

where the\textit{  O(j) }are bounded operators such that \(\sum _j\)\textit{ O}\((\text{\textit{$j$}})^*\)O\textit{ (j)} = \textit{ I}\\

\subsubsection*{ { } { } 4) The channel transforms the sent signal into the received signal. }

 The received signal is now

\[\text{  }\alpha  = \left.\left.\left\langle \delta _z\right.|\rho \left(\delta _z\right.\right)\right\rangle \mathcal{E}\left(G\left(\delta _z\right)\otimes
G\left(\delta _z\right)^*\right)\]

\subsubsection*{\textrm{  { } { } 5) Bob decodes the received signal .}}

He does this by applying \(G^{-1}\) = \(G^*\) { }to the received signal. The state is now\\

\[ G^*\alpha  G = \left.\left.\left\langle \delta _z\right.|\rho \left(\delta _z\right.\right)\right\rangle G^*\circ \mathcal{E}\left(G\left(\delta
_z\right)\otimes G\left(\delta _z\right)^*\right)\circ G,\Pr (\text{received}=\text{sent}|\text{sent}=z)=\left.\left\langle \delta _z| G^*\alpha
 G|\delta _z\right\rangle \right/\text{Trace}(\alpha )\text{                                                          }=\left\langle G\left(\delta
_z\right)\left|\mathcal{E}\left(G\left(\delta _z\right)\otimes G\left(\delta _z\right)^*\right)\right|G\left(\delta _z\right)\right\rangle \]

\subsubsection*{\textrm{  { } { } 6) Bob receives the message. }}

He does this by measuring the positive message observable \(M^{\text{  }}\). Suppose the result is w$\in $FG(2); { }after this the state becomes
\(\left.\left\langle \delta _w|G^*\alpha  G |\right.\delta _w\right\rangle  \delta _w\otimes \delta _w^*\), the probability of this outcome being
\(\frac{\left.\left\langle G\left(\delta _w\right)\right.|\alpha |G\left(\delta _w\right)\right\rangle }{ \left.\left.\left\langle \delta _z\right.|\rho
\left(\delta _z\right.\right)\right\rangle }\). { }

Clearly, Alice and Bob's objective in { }choosing an encoding protocol is to maximize the probability that the message Bob receives is the message
she sends, (before encoding but after sampling from the source $\rho $). { }They use their knowledge of the channel dynamics\textit{  }and [statistical?]
knowledge about the initial channel state and the source $\rho $ { }to design an encoding unitary transformation \textit{ G} { }which maximizes (as
a function of unitary \textit{ G}) the probability that the sent message is the received message.

\begin{equation}\Pr (\text{received}=\text{sent}|\text{source}=\rho ) = \sum _z\Pr (\text{received}=\text{sent}|\text{sent}=z)\Pr (z|\rho )\\
\text{                                                                      }=\sum _z\left\langle G\left(\delta _z\right)\left|\mathcal{E}\left(G\left(\delta
_z\right)\otimes G\left(\delta _z\right)^*\right)\right|G\left(\delta _z\right)\right\rangle \Pr (z|\rho )\end{equation}

 If we assume the operator sum form above then\\
Pr(received=sent$|$source=$\rho $) = \(\left.\left.\sum _{z,j}\Pr (z|\rho )\text{  }|\langle \left.G\left(\delta _z\right)\right|O(j)|G\left(\delta
_z\right)\right\rangle \right|^2\)

Thus Alice and Bob choose G to maximize the above 4\(^{\text{th}}\) degree form in G, subject to the unitary constraint \(G^*\)\textit{ G} = \textit{
I.}

\subsubsection*{\textrm{  { } { } 7) Bob passes the message along to its destination and resets the message state.}}

\textit{ $\texttt{"}$}Passing the message along$\texttt{"}$ { }is a classical process. { }Bob might reset the message state to { }\(\delta _{\Lambda
}\)$\otimes $\(\delta _{\Lambda }^*\) by applying the operator \(R_{w^{-1}}\) . 

\section*{IV. { }HARMONIC { }ANALYSIS ON FG(2)}

A standard reference for harmonic analysis on finitely generated free groups is [TP]. We also use [Na] for some functional analytic aspects of the
argument. The results of this parts are all consequences of results therein. A self-contained development of harmonic analysis of \textit{ FG}(2)\textit{
 }which realizes the principal series on \(L^2\)[0,1[ will be found in [TR3].

The group \textit{ FG}(2) = \(\uplus _{n\geq 0}\) \(\text{FG}(2)_n\) , and so it is countably infinite. With the discrete topology it is a locally
compact group whose Haar (translation-invariant measure) is ordinary summation. We define the spaces \(\ell ^p\)(\textit{ FG}(2)) as usual:

\(\text{Let}\text{  }1 < p < \infty .\text{   }\text{For}\text{  }\text{any}\text{  }\text{complex}-\text{valued}\text{  }\text{function}\text{ 
}\text{\textit{$f$}}\text{   }\text{on}\text{  }\text{\textit{FG}}(2)\text{  }\text{define}\text{  }\text{the}\text{  }\ell ^p\text{  }\text{norm}\text{
  }\text{by}\)

\[\|f\|_p=\left(\sum _{x\in \text{FG}(2)}|f(x)|^p\right)^{1/p}\]

and the space \(\ell ^p\)(FG(2)) , or \(\ell ^p\) { }for short, { }is defined by

\[\ell ^p(\text{FG}(2))=\left\{f : \|f\|_p < \infty \right\}\]

\(\text{If}\text{  }1<p<\infty \text{  }\text{the}\text{  }\text{spaces}\text{  }\ell ^p\text{  }\text{and}\text{  }\ell ^q\text{  }\text{are}\text{
 }\text{dual}\text{  }\text{provided}\\
\text{  }q=p/(p-1).\text{  }\text{The}\text{  }\text{bilinear}\text{  }\text{form}\text{   }\langle \text{\textit{$f$}},\text{\textit{$g$}}\rangle
\text{  }\text{establishes}\text{  }\text{the}\text{  }\text{duality}, \\
\text{where}\)

\begin{equation}\langle f,g\rangle  = \sum _{x\in \text{FG}(2)}\text{\textit{$f$}}(x)\text{\textit{$ $}}\text{\textit{$g$}}(x)\end{equation}

and the sum is absolutely convergent. { }Note this form is invariant under left and rtight translations of f and g together.

Now define

\(\|f\|_{\infty }\) = sup $\{$\texttt{ $|$}f(x)\texttt{ $|$}: x$\in $FG(2)$\}$\\

and the space \(\ell ^{\infty }\) { } as { }the { }space { }of { }all { }functions { }\textit{ f}\textit{  }\textup{  }\textup{ for}\textup{  { }}\textup{
which}\textup{  { }}\(\|f\|_{\infty  }\)\texttt{ $<$} $\infty $ .

Then (11) with \textit{ f $\in $ } \(\ell ^{\infty }\)(\textit{ FG}(2)) and { }\textit{ g$\in $ } \(\ell ^1\)(\textit{ FG}(2))\textit{  { }}identifies
\(\ell ^{\infty }\) with the dual of \(\ell ^1\) { }[but the dual of { }\(\ell ^{\infty }\) { }is { }much larger than { }\(\ell ^1\)].

We are mainly interested in the Hilbert space \(\ell ^2\)(FG(2)), with inner product 

\begin{equation} \langle \text{\textit{$f$}} | \text{\textit{$g$}}\rangle \text{  }=\text{  }\left\langle f^*\right.,\text{  }\text{\textit{$g$}}\text{\textit{$\rangle
$}},\text{  }f^{*\text{\textit{$ $}}}= \text{complex}\text{  }\text{conjugate}\text{  }\text{of}\text{  }\text{\textit{$f$}}\text{\textit{$.$}}\end{equation}

The o.n. basis (\(\delta _x\)) establishes an isomorphism of the QMS $\mathcal{H}$ with  \(\ell ^2\)(\textit{ FG}(2)) connecting f(x) with $\langle
$\(\delta _x\)$\mid $f$\rangle $ . We shall use this identification { }throughout this section. Note \(\delta _x\)(\textit{ y}) equals 1 if \textit{
x}=\textit{ y }and equals 0 otherwise.

\subsection*{A. The Regular Representation\\
}

The \textit{ left and right regular representations L }and \textit{ R} respectively are defined by 

\begin{equation}\text{    }\text{\textit{$L$}}_q\text{\textit{$f$}}(\text{\textit{$x$}}\text{\textit{$)$}}\text{\textit{  }}\text{\textit{$=$}}\text{\textit{$f$}}\left(\text{\textit{$q^{-1}x$}}\text{\textit{$)$}}\right.\text{\textit{
 }}\text{\textit{$,$}}\text{\textit{  }}\text{\textit{$R$}}_q\text{\textit{$f$}}(\text{\textit{$x$}}\text{\textit{$)$}}\text{\textit{  }}\text{\textit{$=$}}\text{\textit{$f$}}(\text{\textit{xq}}\text{\textit{$)$}}\text{
 }\text{where} \text{\textit{$ $}}\text{\textit{$f$}}\text{\textit{$ $}}\text{\textit{$\in $}}\text{\textit{$ $}}\ell ^p(\text{\textit{FG}}(2)) \text{and}\text{
 }\text{\textit{$x$}}\text{\textit{$,$}}\text{\textit{$ $}}\text{\textit{$q$}}\text{  }\text{\textit{$\in $}}\text{\textit{$ $}}\text{\textit{FG}}\text{\textit{$($}}2)
. \end{equation}

The binary operation of \textit{ convolution } * is defined by

\begin{equation}\text{\textit{$f$}}\text{\textit{$ $}}\star \text{\textit{$ $}}\text{\textit{$g$}}\text{\textit{$(x)$}}\text{\textit{$ $}}\text{\textit{$=$}}\text{\textit{$
$}}\sum _{y\cdot z=x}\text{\textit{$f(y)g(z)$}} \end{equation}

\textit{ L }and \textit{ R} define unitary representations of the free group on \(\ell ^2\) \textup{ which { }commute with one another. In fact,
results in [Na]} imply that all operators which commute with \textit{ R} are weak limits of linear combinations of \(L_q\), and conversely. { }More
explicitly, a bounded operator on \(\ell ^2\) { }which { }commutes  with all \(R_q\) { }is of the form { }\(\text{\textit{$g$}}\text{ $|\rightarrow
$}\) h*g { } for some unique \(h\in \ell ^2 \) { }(but not { }every { }such \textit{ h}\textup{  }\textup{ defines}\textup{  { }}\textup{ a}\textup{
 }\textup{ bounded}\textup{  }\textup{ operator}\textup{  { }}\textup{ on}\textup{  }\(\ell ^2\)) . The only operators commuting with both \textit{
L} and \textit{ R} are scalar multiples of \textit{ I}, { }so \textit{ L} (and \textit{ R}) are factor representations of \textit{ FG}(2); in fact,
the factor { }is { }of { }von Neumann type { }\(\text{II}_1\) .

The main connection between QMS and harmonic analysis on \textit{ FG}(2) is the observation that the operators { }\(V_a\) { }which append and remove
bits are the operators { }\(\text{\textit{$R$}}_{a^{-1}}\) .

In the context of harmonic analysis on \textit{ FG}(2) the operators \textit{ k}(\textit{ M}) = \(\sum _{x\in \text{\textit{FG}}(2)} \)k(x) \(\delta
_x\)$\otimes $\(\delta _x^*\) are the multipliers which map \textit{ f }to \textit{ k f. }These operators rarely commute with either \textit{ L }or
\textit{ R. }The measurements which commute with all \(V_a\) commute with \textit{ R }and therefore stem from left convolution operators of the form

\begin{equation}\text{\textit{$f \to  h*f$}}\text{\textit{$ $}}\text{\textit{$=$}}\text{\textit{$ $}} \sum _{x\in \text{\textit{FG}}(2)} h(x)L_x(\text{\textit{$f$}}),
\end{equation}

\textrm{ where { }\( h(x)=h^{\star }(x)=h^*\left(x^{-1}\right.)\) }is such that h*f $\in $ \(\ell ^2\) whenever f $\in $ \(\ell ^2\).

In particular the bit-valued observables \textit{ P }(a.k.a. projectors) which commute with \textit{ R} correspond with functions { }\textit{ p }such
that \(p^{\star }\)$\star $\textit{ p}\textit{  }\textit{ =}\textit{ p}\textit{  }and where \textit{ P}\textup{ (}\textit{ f}\textup{ )}\textup{
 }\textup{ =}\textup{  }\textit{ p}*f . These observables permit conditional operations which are trace-preserving and which commute with all the
operators \(R_a\); i.e., the quantum operations { }$\mathcal{Q}$(\textit{ p}$\star $) and $\mathcal{K}$(\(R_a\)) commute, $\forall $ \textit{ a}$\in
$\textit{ FG}(2). 

\subsection*{B. Radial Functions and the Decomposition { }of { }the Left Regular Representation { }into Irreducible Representations\\
}

The material in this section is adapted from [TP]; a self-contained technical report[TR3] { }is forth-coming. As shown in [Na] { }there is no unique
way to do harmonic analysis on a free group with more than one generator, but the technique in [TP] is elegant and of interest here.

A function \textit{ f }on FG(2) is \textit{ radial }provided \textit{ f}(\textit{ x}) depends only on the message length $\#$\textit{ x . }Let \textit{
K }be the vector space of all radial functions with finite support. The functions (\(\mu _n\)) form an orthogonal basis for \textit{ K, }where \textit{
\(\text{\textit{$\mu _n$}}\)(x)}\textit{ =}\textit{ 1/n}\textit{  { }}\textup{ if}\textup{  }\textit{ $\#$}\textit{ x}\textit{ =}\textit{ n}\textit{
,} and is 0 otherwise.\\
Then \textit{ K} turns out to be a commutative algebra under convolution, and in fact it is generated by \(\mu _1\). This follows from the formula

\begin{equation}\mu _1\star \mu _n = \frac{1}{4}\mu _{n-1}+ \frac{3}{4}\mu _{n+1}, \text{\textit{$n$}}\text{\textit{$>$}}\text{\textit{$0$}} .\end{equation}

It follows immediately by induction that for any \textit{ n }there is a polynomial \(p_n\)($\lambda $)= \(\sum _{k=0}^n \)\(p_{\text{nk}}\)\(\lambda
^k\) such that \(\mu _n\)=\(p_{\text{n0}}\)\(\delta _{\Lambda }\)+ \(p_{\text{n1}}\)\(\mu _1\)+ \(p_{\text{n2} }\)\(\mu _1\)\(\star \mu _1\)+...+
\(p_{\text{nn}}\)\(\mu _1^{\text{$\star $n}}\). In other words, every function in \textit{ K }is a convolution polynomial in \(\mu _1\). The commutative
von Neumann algebra generated by \textit{ A}\textit{  }\textit{ =}\textit{  }\(\mu _1\)$\star $ as an operator on \(\ell ^2\)(\textit{ FG}(2)) is
maximal Abelian in the von Neumann algebra generated by (\(\left.L_a\right)_{a\in \text{FG}(2)}\). This means it can be used to decompose the left
regular representation into irreducible representations of the free group. Note that { }in general the measurement of A will smear a definite message
and scatter its length.

Formula (16) implies that the polynomials \(p_n\) satisfy the recursion

\begin{equation} \left.p_{n+1}\right(\lambda ) = \frac{4}{3} p_n(\lambda ) - \frac{1}{3}p_{n-1}(\lambda ), \text{for}  \text{\textit{$n$}}>0; p_0(\lambda
)=1,\text{  }p_1(\lambda )=\lambda  .\end{equation}

The spectrum of \(\mu _1\)$\star $ { }considered as \textup{ a} self-adjoint operator on \(\ell ^2\)(FG(2)) turns out to be the interval [-\(\frac{\sqrt{3}}{2}\),\(\frac{\sqrt{3}}{2}\)],
so $\lambda $ in (17) may be restricted to that interval. The \(p_n\) are orthogonal polynomials with respect to the weight function \\

\begin{equation}\text{\textit{$w$}}(\lambda )=\frac{1}{\pi }\frac{\sqrt{3-4 \lambda ^2}}{1-\lambda ^2}, |\lambda |\leq \frac{\sqrt{3}}{2}\text{,
} \int _{-\frac{\sqrt{3}}{2}}^{\frac{\sqrt{3}}{2}}p_n(\lambda )p_m(\lambda ) w(\lambda )d\lambda  =\text{  }\frac{(\text{\textit{$n=m$}})}{\#\text{\textit{FG}}(2)_n}
\end{equation}

which equals \(\frac{1}{4}\)\(3^{1-n}\)  if { }0$<$\textit{ m}=\textit{ n} and 0 otherwise.\\
The normalized polynomials \(\epsilon _n\)($\lambda $)=\(\sqrt{\#\text{\textit{FG}}(2)_n}\)\(p_n\)($\lambda $) form an o.n. basis for \(L^2\)([-\(\frac{\sqrt{3}}{2}\),\(\frac{\sqrt{3}}{2}\)],\textit{
w}($\lambda $) $\mathbf{d}\lambda $). The re-scaled polynomials \(E_n\)(x)=\(\epsilon _n\)(\(\frac{\sqrt{3}}{2}\)\textit{ x}), { }$|$\textit{ x$|$}$\leq
$1, satisfy the recursion formula \(E_{n+1}\)(\textit{ x})=2\textit{  x} \(E_n\)(\textit{ x}) - \(E_{n-1}\)(\textit{ x}), which is the recursion
formula satisfied by the Chebyshev polynomials, so \(E_n\) is a linear combination of \(T_n\) and \(U_n\); in fact, for { }n$\geq $1 { }\(E_n\)(x)=(\(T_n\)(x)+\(U_n\)(x))/\(\sqrt{3}\),
so \(p_0\)($\lambda $)=1,

\begin{equation}\forall n>0\text{   }p_n(\lambda )=\frac{1}{2}3^{-n/2}\left(T_n\left(\frac{2}{\sqrt{3}}\lambda \right)+U_n\left(\frac{2}{\sqrt{3}}\lambda
\right)\right)\end{equation}

\subsubsection*{1) The Poisson Boundary of FG(2)\\
}

\textit{ The Poisson Boundary }of the free group is the set $\Omega $ of infinite reduced words from the alphabet 4=$\{$0,1,2,3$\}$. In other words,
{ }$\Omega $ is the subset of the infinite product \(4^{\mathbb{N}}\) consisting of those infinite strings which contain none of the diagrams 02,13,20,31
. $\Omega $ is a closed subset of the compact space \(4^{\mathbb{N}}\) and therefore itself compact.\\
The group \textit{ FG}(2) acts on $\Omega $ in a natural way:

\begin{equation}\text{\textit{$x$}}\cdot \omega  = \text{\textit{reduce}}(x\omega ) = \left.\text{\textit{reduce}}\left(x_0\text{$\ldots $x}_{n-1}\right.\omega
_0\ldots \omega _{n-1}\right)\omega _n\omega _{n+1}.\ldots \end{equation}

E.g., 01121032$\cdot $012231100$\ldots $= 01121231100$\ldots $

For every \textit{ x}$\in $\textit{ FG}(2), \textit{ x}$\neq \Lambda $, { }let \(E_x\)= $\{\omega \in \Omega $: $\forall $j if $\#$\textit{ x$>$j
}then\textit{ , }\(\omega _j\)=\(x_j\)$\}$. There is a unique Borel probability measure $\nu $ on $\Omega $ which assigns measure \(\frac{1}{4\cdot
3^{\#x-1}}\)to \(E_x\). In other words the probability that an infinite reduced word starts with \textit{ x }depends only on the length of \textit{
x.\\
}The action of { }\textit{ FG}(2) on $\Omega $ does not preserve the measure $\nu $; it transforms it according to

\begin{equation}\text{   }\int _{\Omega } \xi (x\cdot \omega )d\nu  = \int _{\Omega } P(x,\omega )\xi (\omega )d\nu , \text{for}\text{  }\text{all}\text{
 }\text{continuous}\text{  }\text{functions}\text{  }\xi \text{  }\text{on}\text{  }\Omega .\end{equation}

In [TR3] an elementary measure isomorphism between ($\Omega $,$\nu $) and [0,1] with Lebesgue measure is established so in fact all the analysis
may be transferred to the unit interval.

\subsection*{2) The Poisson Kernel and the Principal Series}

\textit{ The Poisson Kernel }is the function \textit{ P }in the above formula (21); it is given by P(x,$\omega $)=\(3^{2N(x,\omega )-\#x}\), where
N(x,$\omega $)=\textit{ k }iff  \(\omega _j\)=\(x_j\) { }if \textit{ k} $>$ \textit{ j { }}but \(\omega _k\)$\neq $\(x_k\). This formula { }implies
that the operator \(\pi _z\)(x) on \(L^2\)($\Omega $,$\nu $) is unitary whenever { } { } Re(\textit{ z})=\(\frac{1}{2}\):

\begin{equation} \pi _{1/2 + i t}(x)(\xi )(\omega )=P(x,\omega )^{1/2+i t}\xi \left(x^{-1}\right.\cdot \omega ),\end{equation}

\textrm{  }\textrm{ where}\textrm{  }\textrm{  }\textrm{ \textit{ t}}\textrm{ \textit{ $\in $}}\textrm{ $\mathbb{R}$}\textrm{  }\textrm{ and}\textrm{
 }\textrm{ $\xi $$\in $\(L^2\)}\textrm{ ($\Omega $,$\nu $)}. In fact, \textit{ x}$\to $\(\pi _{1/2 + i t}\)(x) defines a unitary representation of
the free group \textit{ FG}(2) on \(L^2\)($\Omega $,$\nu $), one of the members of the \textit{ principal series }of representaions \(\pi _z\) .
When z = 1/2+i t the representation is \textit{ irreducible }in that no non-trivial closed Hilbert subspace of \(L^2\)($\Omega $,$\nu $) is invariant
under all \(\pi _z\)(x). Irreducibility implies that all bounded linear operators on \(L^2\)($\Omega $,$\nu $) are limits (in the weak operator topology)
of { }finite linear combinations of { }\(\pi _{1/2 + i t}\)(x), and that all pure states are limits of linear combinations of { }\(\pi _{1/2 + i
t}\)(x)($\varphi $), where $\varphi $ is any non-zero vector (\(1_{\Omega }\) say). 

\subsubsection*{3) The Fourier Transform and the Decomposition of the Left Regular Representation\\
}

The\textit{  Fourier Transform }which decomposes the left regular representation can now be defined. Let

\begin{equation} \text{\textit{$K$}}(\text{\textit{$($}}\lambda ,\omega ),\text{\textit{$x$}}) = P(x,\omega )^{1/2+i \mathit{t}(\lambda )},\end{equation}

where\textit{  }\textrm{  }\textrm{ \textit{ x}}\textrm{ $\in $}\textrm{ \textit{ FG}}\textrm{ (2)}\textrm{ ,}\textrm{  }\textrm{ $\omega $$\in $$\Omega
$}\textrm{ ,}\textrm{  }\textrm{ $\lambda $}\textrm{ $\in $}\textrm{  }sp(A)=[-\(\frac{\sqrt{3}}{2}\),\(\frac{\sqrt{3}}{2}\)]\textrm{  }\textrm{
,}\textrm{  }\textrm{ and}\textrm{  }\( \mathit{t}(\lambda )=\frac{\cos ^{-1}\left(\frac{2}{\sqrt{3}}\lambda \right)}{\ln (3)}\)\textrm{ \\
}.\\
.\textit{ K }is the kernel for the Fourier transform:

\begin{equation}\text{      }\mathcal{F}(\text{\textit{$f$}})(\lambda ,\omega ) = \sum _{x\in \text{FG}(2)} \text{\textit{$K$}}(\text{\textit{$($}}\lambda
,\omega ),\text{\textit{$x$}})f(x)\end{equation}

\[\mathcal{F}:\left.\ell ^2\right(\text{\textit{FG}}(2)) :\to  L^2(\text{sp}(A)\times \Omega ,w(\lambda )\mathbf{d} \lambda \otimes d\nu ) , .\text{
                 }\]

\(\text{  }\text{                    }\delta _{\Lambda } \to  1_{\text{sp}(A)\times \Omega }\\
\text{                      }\delta _y \to  \text{\textit{$K$}}(\text{\textit{$($}}\lambda ,\omega ),\text{\textit{$y$}})\\
\text{                       }\mu _n\to  p_{n\otimes 1_{\Omega }}\)

$\mathcal{F}$ decomposes left convolution operators:

\( \mathcal{F}(\text{\textit{$g* f$}}\text{\textit{$)$}}\text{\textit{$($}}\lambda ,\omega ) = \sum _y g(y) \pi _{1/2+\text{it}(\lambda )}(y)(\mathcal{F}(f)(\lambda
,.))(\omega ) , \\
 \text{or}\text{  }\text{equivalently} \\
\mathcal{F}\circ (\text{\textit{$g*$}}\text{\textit{$)$}}\text{\textit{$ $}}\text{\textit{$\circ $}}\text{\textit{$ $}}\mathcal{F}^{-1}=\int _{\text{sp}(A)}
\sum _y g(y)\pi _{1/2+\text{it}(\lambda )}(y) w(\lambda )d\lambda  \text{  }\)

expressing the Fourier-transformed operator as { }a $\texttt{"}$direct integral$\texttt{"}$ . In particular,

\[ \mathcal{F}\circ L_a\circ \mathcal{F}^{-1}=\int _{\text{sp}(A)}\pi _{1/2+i \mathit{t}(\lambda )}(a)w(\lambda )d\lambda \]

decomposes the left regular representation as a direct integral of irreducible representations of the principal series.

Recall that { }A = left convolution by \(\mu _1\)=(\(\delta _0\)+\(\delta _1\)+\(\delta _2\)+\(\delta _3\))/4 . Then $\mathcal{F}$ diagonalizes all
the operators \textit{ k}(\textit{ A}), where \textit{ k }is any bounded Borel function: $\mathcal{F}$(\textit{ k}(\textit{ A})(\textit{ f}))($\lambda
$,$\omega $) = \textit{ k}($\lambda $) $\mathcal{F}$(\textit{ f})($\lambda $,$\omega $),\\
or equivalently $\mathcal{F}$$\circ $k(A)$\circ $\(\mathcal{F}^{-1}\)=Mul(\textit{ k}$\otimes $\(1_{\Omega }\)). { }In particular \textit{ A} is
transformed to multiplication by $\lambda $ on [-\(\sqrt{3}\)/2,\(\sqrt{3}\)/2] .\\
The inverse Fourier transform is given by the kernel { }:

\(K^{\top }((\lambda ,\omega ),\text{\textit{$x$}}\text{\textit{$)$}}\text{\textit{$ $}}\text{\textit{$=$}}\text{\textit{$ $}}K(x,(\lambda ,\omega
))^*= P(x,\omega )^{1/2-i \mathit{t}(\lambda )}\\
\mathcal{F}^{-1}(\Psi )(\text{\textit{$x$}}\text{\textit{$)$}}\text{\textit{$ $}}\text{\textit{$=$}}\text{\textit{$ $}}\int _{\text{sp}(A)}\int _{\Omega
} \left.K^{\top }\right((\lambda ,\omega ),\text{\textit{$x$}}\text{\textit{$)$}}\Psi (\lambda ,\omega )d\nu (\omega ) w(\lambda )d\lambda .\)

Thus the transform is unitary; here is the Plancherel formula:

\[ \|f\|_2^2\text{\textit{$ $}}= f^*\text{\textit{$*$}}\text{\textit{$ $}}\text{\textit{$f$}}\text{\textit{$ $}}(\Lambda ) = \int _{\text{sp}(A)}\int
_{\Omega } |\mathcal{F}(f)(\lambda ,\omega )|^2d\nu (\omega ) w(\lambda )d\lambda \]

This summary of harmonic analysis on the free group \textit{ FG}(2) indicates that one class of projectors what commute with all \(R_a\) are those
of the form \\
\textit{ f} $\to $ \textit{ g }$\star $ \textit{ f },where \textit{ g} is a radial function. The Fourier transform of such an operator is of the
form Mul(\(\hat{g}\)$\otimes $\(1_{\Omega }\)), where $\mathcal{F}$(\textit{ g)=}\(\hat{g}\)$\otimes $\(1_{\Omega }\)\textit{  . }The operator is
a projector iff \(\hat{g}^2\) = \(\hat{g}\), so \(\hat{g}\) = \(1_B\) for some Borel set B $\subseteq $ sp(A) . If follows that

\begin{equation}\text{  }\text{\textit{$g$}}(\text{\textit{$x$}}\text{) }=G_n(\text{\textit{$B$}}\text{)= }\int _B\int _{\Omega }P\left(x^{-1},\omega
\right)d\nu (\omega )w(\lambda )d\lambda \text{ = }\int _Bp_n(\lambda )w(\lambda )d\lambda  \end{equation}

where \textit{ n }= $\#$\textit{ x, g} is a radial function, and \textit{ g}($\Lambda $) = \(\int _B\)w($\lambda $)d$\lambda $.  Since \(\text{\textit{$g$}}\star
\delta _{\Lambda }\) = \textit{ g}, { }for all \textit{ x}$\in $FG(2) $|$g(x)\(|^2\)/($\|$g$\|$\(|_2^2\)) represents the probability of receiving
message \textit{ x }after measuring the projector \textit{ g}$\star $ starting in state \(\delta _{\Lambda }\). This number is the same for all reduced
words of the same length \textit{ n}, { }given by ( \(\int _B\)\(p_n\)($\lambda $)w($\lambda $)d$\lambda $ \()^2\). Note that this probability is
not an additive set function of \textit{ B, }whereas\textit{  B$\to $}\(G_n\)(\textit{ B}) is a signed measure.

\subsection*{C. The Harmonic Analysis of a Single { }Left Translate Operator \(L_a\)}

Let \textit{ a }$\in $ \textit{ FG}(2), \textit{ a} $\neq $ $\Lambda $. Such a choice defin$\mathcal{F}$es a non-trivial action of the integers $\mathbb{Z}$
on \textit{ FG}(2) : (\textit{ n},\textit{ x}) $\to $ \(a^n\)$\cdot $ \textit{ x . }Each orbit has a unique reduced word \textit{ z }of minimum length.
Therefore, if \textit{ z}$\neq \Lambda $, \(z_0\)$\neq $\textit{ inv}\textup{ (}\(a_{\#a-1}\))= inverse of last letter of \textit{ a, }and \(z_0\)$\neq
$\(a_0\) = first letter of { }\textit{ a. }Therefore the orbits are in one-to-one correspondence with the set \textit{ Z }of all \textit{ z} satisfying
these conditions. Thus there is a unique unitary operator $\mathcal{T}$:\(\ell ^2\)(FG(2)) $:\to $\(L^2\)($\mathbb{T}\times $\textit{ Z}), where
$\mathbb{T}$ is the unit circle with normalized euclidean measure { }\(\frac{d\theta }{2\pi }\), such that $\mathcal{T}$(\(\delta _{a^n\cdot z}\))($\alpha
$,z) = \(\alpha ^n\)\(\delta _z\) . This operator diagonalizes \(L_a\); i.e., $\mathcal{T}$ $\circ $\(L_a\)$\circ $\(\mathcal{T}^{-1}\)= { }Mul($\zeta
$$\otimes $\(1_Z\)), { }$\zeta $($\alpha $)=$\alpha $, $\mathcal{T}$ $\circ $\(L_a\)$\circ $\(\mathcal{T}^{-1}\)(\(\sum _{n\in \mathbb{Z}} \)\(\hat{k}\)(\textit{
n}) \(L_{a^n}\)) = Mul(k$\otimes $\(1_Z\)) for any bounded function \textit{ k }on $\mathbb{T}$ with ordinary Fourier series \(\hat{k}\)(\textit{
n}) = \(\frac{1}{2\pi }\)\(\int _{-\pi }^{\pi }\) k(\(e^{\text{i$\theta $}}\))\(e^{-i \text{n$\theta $}}\)d$\theta $, and the operator \(\sum _{n\in
\mathbb{Z}} \)\(\hat{k}\)(\textit{ n}) \(L_{a^n}\) is just convolution with g=\(\sum _{n\in \mathbb{Z}} \)\(\hat{k}\)(\textit{ n}) \(\delta _{a^n}\).
The functions \textit{ k }which yield projectors in the weakly closed algebra generated by \(L_a\)and \(L_a^*\) are indicator functions of Borel
subsets of $\mathbb{T}$. 

The case \textit{ k }= \(1_{\{\alpha : |\arg(\alpha )|<\phi \}}\), $\pi $$\succeq $$\phi $$>$0, determine projectors which are easy to compute, since

\[ g\left(a^n\right)= \hat{k}(\text{\textit{$n$}}) = \frac{1}{2\pi }\int _{-\pi }^{\pi } k\left(e^{\text{i$\theta $}}\right)e^{-i \text{n$\theta
$}}d\theta  = \frac{1}{2\pi }\int _{-\phi }^{\phi } e^{-i \text{n$\theta $}}d\theta  = \sin (\text{\textit{$n$}}\phi )/(\text{\textit{$n$}}\pi )
\]

 if n$\neq $0, and \(\hat{k}\)(0) = \(\frac{\phi }{\pi }\) = \(\|g\|_2^2\) since \(k^2\) = \textit{ k }in this case. \\
Thus \(L_a\) is unitarily equivalent to an infinite sum of bilateral shifts . Of course \(L_a\) and all the operators \(\sum _{n\in \mathbb{Z}} \)\(\hat{k}\)(\textit{
n}) \(L_{a^n}\) commute with all the operators \(R_y\). Finally, since \(R_a\)= V$\circ $\(L_a\)$\circ $\(V^*\), where \textit{ V} is the unitary
reflexion operator V(f)(x)=f(\(x^{-1}\)), \(R_a\) is also unitarily equivalent to an infinite sum of bilateral shifts.

\section*{V. { }DISCUSSION { }AND { }CONCLUSIONS}

In this report we have developed a mathematical formulation of quantum message space and demonstrated a type of calculus for incorporating simple
operations on bits into quantum communication theory and computation. How might quantum message space be implemented physically? Note that since
a quantum message space is basically a Hilbert space with a complete orthonormal basis indexed by \textit{ FG}(2), the free group on two elements,
the problem is $\texttt{"}$merely$\texttt{"}$ one of labelling. For example, the energy levels of a harmonic oscillator with one degree of freedom
are of the form a n + b, where n = 0,1,2,3,... . One could employ any method for listing the elements of FG(2) to index the eigenstates of the oscillator
by FG(2). Of course, in this case since the multiplicity is fixed the amount of energy associated with a message grows exponentially with the message
length.\\
Or consider a hydrogen atom. The bound states are of the form { }a/n + b , each with multiplicity 2 n + 1 . In principle, FG(2) could label the multitude
of energy stationary states. But now the energy levels pile up around 0, and the practical problems of distinguishing bound states when n is very
large would make it difficult to send long messages.\\
In their classic text, Nielsen and Chaung [NC,p.203] speculate:\\

$\texttt{"}$...it is by no means clear that the basic assumptions underlying the [finite dimensional] state space and { }starting { }conditions [from
the computational basis] { }in the quantum circuit model are justified. Might there be anything { }to be gained by using { }systems whose state space
is infinite dimensional? Assuming that the starting state...is a computational basis state is also not necessary...$\texttt{"}$ 

They go on to suggest that other states, other basic unitary processors, and other measurements might be able to $\texttt{"}$...perform tasks intractible
within the quantum circuit model.$\texttt{"}$ We doubt if QMS and the operators we've studied extend the realm of theoretical computability, but
they do show how a consistent quantum model of message sources of arbitrary length leads naturally to considerations of alternate models.

\section*{APPENDIX: Proof That { }The Source Entropy { }$\geq $ Von Neumann Entropy}

Let $\sigma $ be any normalized state, and let (\(u_k\)) be an orthonormal basis of eigenvectors for $\sigma $=\(\sum _k \)\(\sigma _k\)\(u_k\)$\otimes
$\(u_k^*\),\(\underset{k}{\text{    }\sum }\)\(\sigma _k\)=1 $\forall $\textit{ x}\textit{  }$\in $ \textit{ FG(2)}, $\langle $\(\delta _x\) $\mid
$$\sigma $(\(\delta _x\))$\rangle $=\(\sum _k \)\(\sigma _k\)\(\left|\text{\textit{$u_{\text{kx}}$}}\text{\textit{$|$}}\right.^2\), { }where { }\textit{
\(\text{\textit{$u_{\text{kx}}$}}\)}=$\langle $\(\delta _x\)$\mid $\(u_k\)$\rangle $. Since the function $\eta $ { }is convex downward and { }\(\sum
_k |\text{\textit{$u_{\text{kx}}$}}|^2=\left|\delta _x\right.|^2 =1\), { }$\eta $($\langle $\(\delta _x\) $\mid $$\sigma $(\(\delta _x\))$\rangle
$) $\geq $ \(\sum _k \)$\eta $(\(\sigma _k\))\(\left|\text{\textit{$u_{\text{kx}}$}}\text{\textit{$|$}}\right.^2\). { }Now sum on { }\textit{ x,
}and we see 

\(\text{\textit{source}}\text{\textit{  }}\text{\textit{entropy}}(\sigma ) \geq  \sum _x  \sum _k \eta \left(\sigma _k\right)|\text{\textit{$u_{\text{kx}}$}}|^2=
\sum _k \eta \left(\sigma _k\right)\sum _x |\text{\textit{$u_{\text{kx}}$}}|^2 =\\
\sum _k \eta \left(\sigma _k\right)=\text{Trace}(\eta (\sigma ))= \text{von}\text{  }\text{Neumann}\text{  }\text{entropy}(\sigma ).\)

\section*{References}

[BEZ] D. Bouwmeester, Artur Ekert, Anton Zeilinger (Eds.), \textit{ The Physics of Quantum Information}. New York: Springer-Verlag, 2000.

[Gr] { }H. S. Green, \textit{ Information Theory and Quantum Physics}. New York: Springer-Verlag, 2000 .

[KB] { }S. Kullback, \textit{ Information Theory and Statistics.} New York: Dover Publications, 1968.

[Ma] G. Mackey, \textit{ Mathematical Foundations of Quantum Mechanics.}New York: W. A. Benjamin, Inc., 1963 .

[Na] { }M.A. Naimark, \textit{ Normed Rings.} Groningen, The Netherlands: P Noordhoff, Ltd., 1964 

[NG] I. Nielsen and Isaac Chaung, \textit{ Quantum Computation and Quantum Information. }New York: Cambridge University Press, 2002 .

[PS] { }Michael E. Peskin, \textit{ Introduction to Quantum Field Theory. }New York:\textit{  }HarperCollins, 1995.

[TP] A. Figa-Talamanca and M. Picardello, \textit{ Harmonic Analysis on Free Groups}. New York: Marcel-Dekker, Inc., 1983.

[TR3] R. D. Ogden, \textit{ Harmonic Analysis in the Free Group with Two Generators}.Technical Report $\#$3, Computer Science Department, Texas State
University at San Marcos, to appear.

\end{document}